\documentclass[aps,pra,10pt,twocolumn,linenumbers,groupedaddress]{revtex4-1}

\usepackage{hyperref}
\usepackage{amsmath}
\usepackage{amsfonts}
\usepackage{amsbsy}
\usepackage{amssymb}
\usepackage{latexsym}
\usepackage{subfigure}
\usepackage[pdftex]{graphicx}
\usepackage{verbatim}
\usepackage{epstopdf}

\newcommand{\yb}{^{171}\text{Yb}^{+}}

\newcommand{\lpl}{\text{lin}\perp\text{lin}}
\newcommand{\sech}{\,\text{sech}}
\newcommand{\nth}{n^\text{th}}

\newcommand{\ket}[1]{\left | #1 \right \rangle}

\newcommand{\oq}{\omega_\text{q}}
\newcommand{\otrap}{\omega_\text{trap}}
\newcommand{\oa}{\omega_\text{A}}
\newcommand{\orep}{\omega_\text{rep}}

\newcommand{\tp}{\tau}

\newcommand{\sx}{\hat{\sigma}_x}

\newcommand{\sz}{\hat{\sigma}_z}
\newcommand{\splus}{\hat{\sigma}_{+}}
\newcommand{\sminus}{\hat{\sigma}_{-}}
\newcommand{\spm}{\hat{\sigma}_{\pm}}
\newcommand{\smp}{\hat{\sigma}_{\mp}}
\newcommand{\ufe}{U_\text{FE}}

\graphicspath{{./Figures/}}

\begin{document}

%Title of paper
\title{Quantum Control of Qubits and Atomic Motion Using Ultrafast Laser Pulses}

\author{J. Mizrahi}
\email{mizrahi.jonathan@gmail.com}
\author{B. Neyenhuis}
\author{K. Johnson}
\author{W. C. Campbell}
\author{C. Senko}
\author{D. Hayes}
\author{C. Monroe}
\affiliation{Joint Quantum Institute, University of Maryland Department of Physics and National Institute of Standards and Technology,
College Park, Maryland 20742}

\date{\today}

\begin{abstract}
Pulsed lasers offer significant advantages over CW lasers in the coherent control of qubits.  Here we review the theoretical and experimental aspects of controlling the internal and external states of individual trapped atoms with pulse trains.  Two distinct regimes of laser intensity are identified.  When the pulses are sufficiently weak that the Rabi frequency $\Omega$ is much smaller than the trap frequency $\otrap$, sideband transitions can be addressed and atom-atom entanglement can be accomplished in much the same way as with CW lasers.  By contrast, if the pulses are very strong ($\Omega \gg \otrap$), impulsive spin-dependent kicks can be combined to create entangling gates which are much faster than a trap period.  These fast entangling gates should work outside of the Lamb-Dicke regime and be insensitive to thermal atomic motion.
\end{abstract}

\maketitle

\section{\label{sec:intro}Introduction}

Over the past decade, frequency combs from mode-locked lasers have become an essential tool in the field of optical frequency metrology \cite{Udem,Cundiff,Hall,Hansch}.  This is due to the broad spectrum of lines spaced by the pulse repetition rate present in a frequency comb.  This allows it to serve as a precise connection between distant frequencies.  In the context of metrology, this feature is used as a ruler in which the spacings between comb lines serve as tick marks.  In the context of coherent control, this feature can be used to directly bridge large frequency gaps between energy levels in a controllable way.  This technique has been used effectively to control diverse quantum systems, including multilevel atoms \cite{Stowe}, molecules \cite{Viteau}, semiconductor spin states \cite{Press, Greilich}, and ions \cite{Hayes, Campbell, Mizrahi}.  Mode-locked lasers therefore have a bright future as a tool for qubit manipulation in a number of different quantum computer architectures.

Trapped atomic ions are a very promising medium for quantum information, due to their extremely long coherence times, well-established means for coherent control and manipulation, and potential for scalability\cite{BlattWinelandReview, Ladd}.  High fidelity entanglement of ions is now routinely achieved\cite{Sackett,Haffner,Haljan,Monz}, as well as implementations of schemes for analog quantum simulation\cite{Islam,Barreiro,Friedenauer} and digital quantum algorithms\cite{Gulde,Brickman,Schindler}.  However, obstacles remain before a trapped ion quantum computer can outperform a classical computer.  Technical limitations to gate fidelity include laser induced decoherence \cite{Ozeri,Ozeri2} and ion heating \cite{Turchette}.  Existing gates are also typically limited in the number of ions which can be manipulated in a single chain. This is because these gates rely on addressing normal modes of motion of the ion chain\cite{Molmer,Milburn}.  As the number of ions grows, the density of normal modes in frequency space grows as well, making it increasingly difficult to avoid undesired couplings. This increased mode density slows down the gate, increasing sensitivity to low frequency noise.

High power mode-locked lasers offer one potential solution to some of these issues (there are a number of other approaches, see \cite{Hensinger,Home,Ospelkaus,Olmschenk2}). The goal of this paper is to discuss recent work on the interaction between trapped ions and mode-locked laser pulses.

From a technical standpoint, the large bandwidth inherent in a comb eliminates some of the complexity and expense of driving Raman transitions.  For hyperfine qubits in ions, the frequency splitting is typically several GHz.  Bridging this gap with CW beams requires either two separate phase-locked lasers, or a high frequency EOM (which is typically inefficient).  By contrast, a single mode-locked laser can directly drive the transition without any high frequency shifts.  Moreover, it is not necessary to stabilize either the carrier-envelope phase or the repetition rate of the mode-locked laser, as will be discussed later.  This enables the use of commercially available, industrial lasers.

As a second advantage, the large instantaneous intensity present in a single pulse enables efficient harmonic generation.  For this reason, high power UV lasers are readily obtainable at frequencies appropriate for trapped ion control. High power enables operating with a large detuning, which suppresses laser-induced decoherence. High power also enables motion control in a time significantly faster than the trap period, which is a new regime in trapped ion control. It should allow the implementation of theoretical proposals for ultrafast gates which are independent of ion temperature, as discussed in section \ref{sec:Gates}.

This paper is divided into three parts. Section \ref{sec:SpinControl} describes spin control of an ion with a pulse train, without motional coupling. Section \ref{sec:SpinMotion} introduces spin-motion coupling. Section \ref{sec:Gates} explains how to realize an ultrafast two ion gate using fast pulses.

\subsection{Experimental System}
We take the atomic qubit as composed of stable ground state electronic levels separated by rf or microwave frequencies.  The schemes reported here can be extended to the case of qubit levels separated by optical intervals, but for concreteness we will concentrate on qubits stored in hyperfine or Zeeman levels in the ground state of an alkali-like atom.

In order to effectively use laser pulses for qubit control, we require three frequency scales to be well separated.  Let $\tp$ denote the pulse duration. The pulse bandwidth 
$1/\tp$ should be much larger than the qubit splitting $\oq$ so that the two qubit levels are coupled by the optical field, yet it should be much smaller than the detuning $\Delta$ from the excited state so that it is negligibly populated during the interaction.  Note also that the detuning $\Delta$ should not be much larger than the fine structure splitting in an alkali-like atom, otherwise the Raman coupling is suppressed \cite{Haljan}.  For many atomic systems, the condition $\oq \ll 1/\tp \ll \Delta$ is satisfied for a range of laser pulse durations $0.5$ ps $\lesssim \tp \lesssim 25$ ps.

Here we consider the interaction between ultrafast laser pulses and qubits represented by laser-cooled $\yb$ ions confined in an RF Paul trap, although many of the results discussed herein are applicable in a range of contexts involving ultrafast pulses on the internal and external degrees of freedom of optically-coupled qubits.  The qubit levels are defined by the $m_F=0$ states of the $^2S_{1/2}$ hyperfine manifold of $\yb$: $\ket{F=0, m_F=0}\equiv\ket{0}$, $\ket{F=1, m_F=0}\equiv\ket{1}$. The qubit frequency splitting is $\oq/2\pi = 12.6\:\text{GHz}$. Doppler cooling of atomic motion, and initialization/detection of the qubit are all accomplished using continuous wave (CW) beams near 369 nm \cite{Olmschenk}. 
\begin{figure}%
\includegraphics[width=\columnwidth]{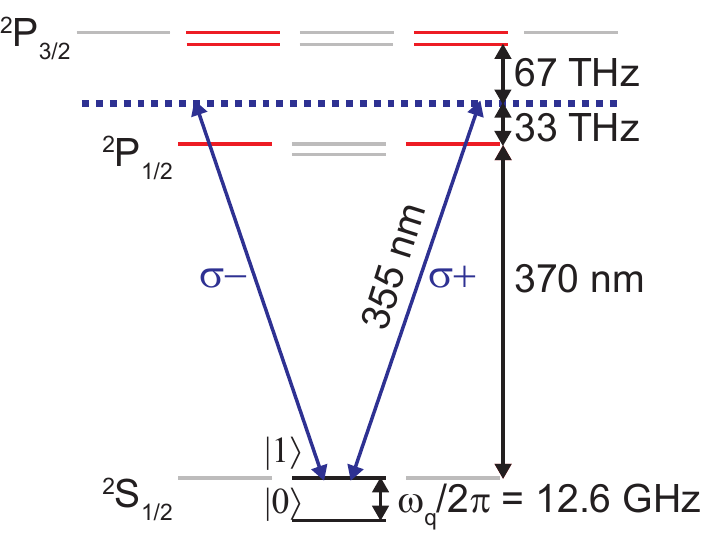}%
\caption{Relevant energy levels of $\yb$. The qubit is identified with the two $m_F=0$ states in the ground state manifold. Continuous wave 369 nm light is used for cooling, detection, and optical pumping.  Laser pulses at 355 nm are used for qubit manipulation, driving stimulated Raman transitions between the qubit levels from $\sigma_{\pm}$ polarized light.}%
\label{fig:YbLevels}%
\end{figure}

\begin{figure}%
\includegraphics[width=\columnwidth]{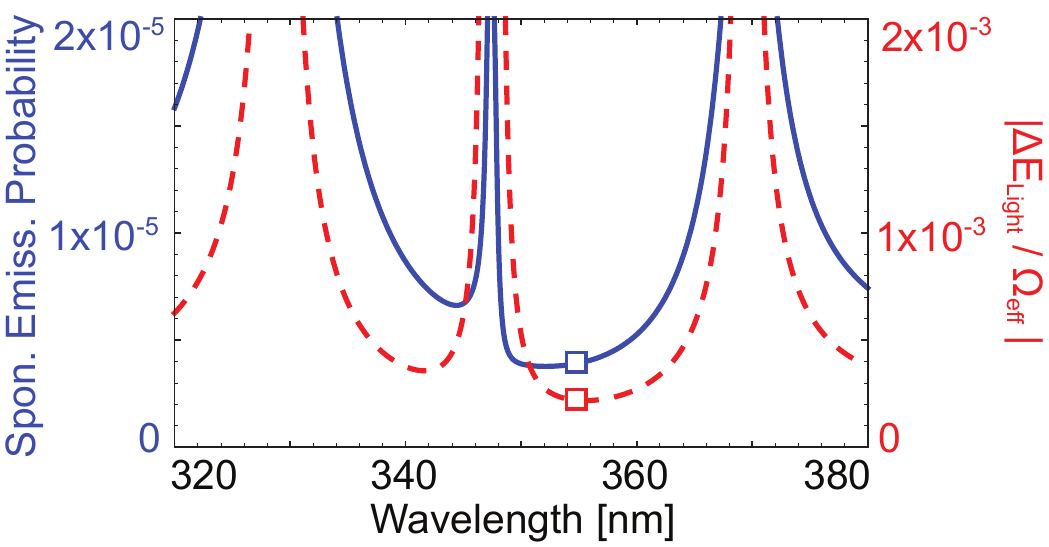}%
\caption{Theoretical curves showing sources of laser-induced decoherence and Stark shifts as a function of wavelength. Solid blue line is spontaneous emission probability during a $\pi$ pulse as a function of laser wavelength.  Dashed red line is differential AC Stark shift divided by Rabi frequency as a function of laser wavelength.  White squares are at 355 nm, where both curves are near a minimum.}%
\label{fig:LightShiftSE}%
\end{figure}

We consider optical pulses generated from a mode-locked tripled Nd:YVO$_4$ laser at 355 nm to drive stimulated Raman transitions between the qubit states $\ket{0}$ and $\ket{1}$, that may also be accompanied by optical dipole forces.  Typical laser repetition rates are in the range $\orep/2\pi = 80-120$ MHz, with a pulse duration $\tp \sim$10 ps ($\sim$100 GHz bandwidth) and maximum average power $\bar{P}$ of several Watts (pulse energies of up to $100$ nJ).  This light is detuned by $\Delta_{1/2}\approx +33$ THz from the excited $^2P_{1/2}$ level, and $\Delta_{3/2}\approx -67$ THz from the $^2P_{3/2}$ level, as shown in figure \ref{fig:YbLevels}. This wavelength and pulse duration is nearly optimal for controlling the $\yb$ system, exhibiting minimal spontaneous emission and differential AC Stark shifts \cite{Campbell}, as shown in figure \ref{fig:LightShiftSE}.

\section{\label{sec:SpinControl}Spin Control with Pulses}
\subsection{\label{sec:StrongPulseSpin}Strong Pulses}
\begin{figure}%
\includegraphics[width=\columnwidth]{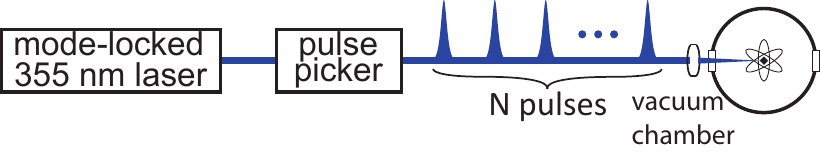}%
\caption{A fast pulse picker selects a train of $N$ circularly polarized pulses, each with area $\theta$. These pulses drive 
simulated Raman transitions in a trapped ion.}%
\label{fig:SimpleExperiment}%
\end{figure}

Consider the interaction of a train of pulses with an atom, as shown in figure \ref{fig:SimpleExperiment}. 
After performing a rotating wave approximation at the optical frequency and adiabatically eliminating the excited $P$ states, the effective Hamilitonian for the interaction becomes \cite{Campbell}:
\begin{equation}
\mathit{H} = -\frac{\oq}{2}\hat{\sigma}_z-\frac{\Omega(t)}{2}\hat{\sigma}_x
\label{eq:Hamiltonian}
\end{equation}
where $\oq$ is the qubit splitting, $\hat{\sigma}_{z,x}$ are Pauli spin operators, and $\Omega(t) =g(t)^2/2\Delta$ is the two-photon Rabi frequency for pure $\sigma^+$ or $\sigma^-$ polarized light. Here, the single-photon $S-P$ resonant coupling strength $g(t)=\gamma\sqrt{I(t)/2I_{sat}}$ with effective detuning given by $1/\Delta = 1/\Delta_{1/2} - 1/\Delta_{3/2}$, accounting for both excited states.  $I(t)$ is time-dependent intensity of the pulse. In the $\yb$ system, $I_{sat}=0.15$ W/cm$^2$ is the saturation intensity for the $^2S_{1/2}-^2P_{1/2}$ transition and the $^2P_{1/2}$ state linewidth is $\gamma/2\pi = 19.6$ MHz.  

We note that the above Hamiltonian can be generalized to include the effect of ultrafast pulses connecting the qubit levels to a third (transiently populated) level on resonance, or in the case of qubits with an optical splitting, directly on resonance with the qubit levels \cite{Poyatos}.  In addition, by choosing appropriate qubit levels and laser pulse polarization, a generalization of the above interaction can produce a differential Stark shift instead of a transition between the levels, in which case the $\hat{\sigma}_{x}$ coupling term above is replaced by $\hat{\sigma}_{z}$ \cite{Poyatos}.  In this case, the actual implementation of entangling gates between multiple qubits through collective motion is not exactly as described below, although there are many similarities.  It should also be noted that qubit states that have sizable differential AC Stark shift are also first-order sensitive to external magnetic fields \cite{Haljan}, and hence perform as relatively poor qubit memories.  

For a single pulse ($N=1$) with either $\sigma_{\pm}$ polarization, the time dependence of the Rabi frequency originates from the intensity profile of the laser $I(t)$, which for a mode-locked laser pulse can be accurately modelled by a squared hyperbolic secant envelope \cite{Siegman}.  Intensity envelope functions of externally generated optical harmonics of the fundamental laser field should be higher powers of the sech function. However, their shape remains quite similar to that of the sech function. We therefore approximate the pulse intensity as 
$I(t)=I_0 \sech\left(\frac{\pi t}{\tp}\right)$ with peak laser intensity $I_0$ and pulse duration $\tp$, having FWHM in time of 
$0.838\tp$.  This approximation allows a simple analytic solution to the evolution of the above Hamiltonian, and numerical simulation indicates that this is at most a 1-2\% correction to everything presented here.  

The qubit Rabi frequency can therefore be written as:
\begin{equation}
\Omega(t) = \frac{\theta}{\tp}\sech\left(\frac{\pi t}{\tp}\right),
\label{eq:OnePulseRabiFreq}
\end{equation}
where $\theta = \int{\Omega(t)dt}$ is the pulse area. For the Raman transition considered here in the 
$\yb$ system using light tuned to $355$ nm, we have \cite{Hayes}:
\begin{equation}
\theta = \frac{I_0\tp\gamma^2}{2I_{sat}\Delta} \label{eq:theta}
\end{equation}
Alternatively, $\theta$ can be expressed in terms of the average intensity of the laser $\bar{I}$ and the repetition frequency $\orep$ using the relation $I_0\tp=2\pi\bar{I}/\orep$.  We find that to drive a Raman $\pi$-pulse with a single laser pulse focussed to a Gaussian waist $w$ (1/e field radius), the required pulse energy is
$\mathcal{E}_{\pi} = \pi I_0 w^2 \tp/2 =   \pi^2 I_{sat}w^2 \Delta/\gamma^2$.  For the $\yb$ system using a 355 nm beam focused to a waist of $w=10$ $\mu$m, we find $\mathcal{E}_{\pi} \sim 12$ nJ. 

\begin{figure}%
\includegraphics[width=\columnwidth]{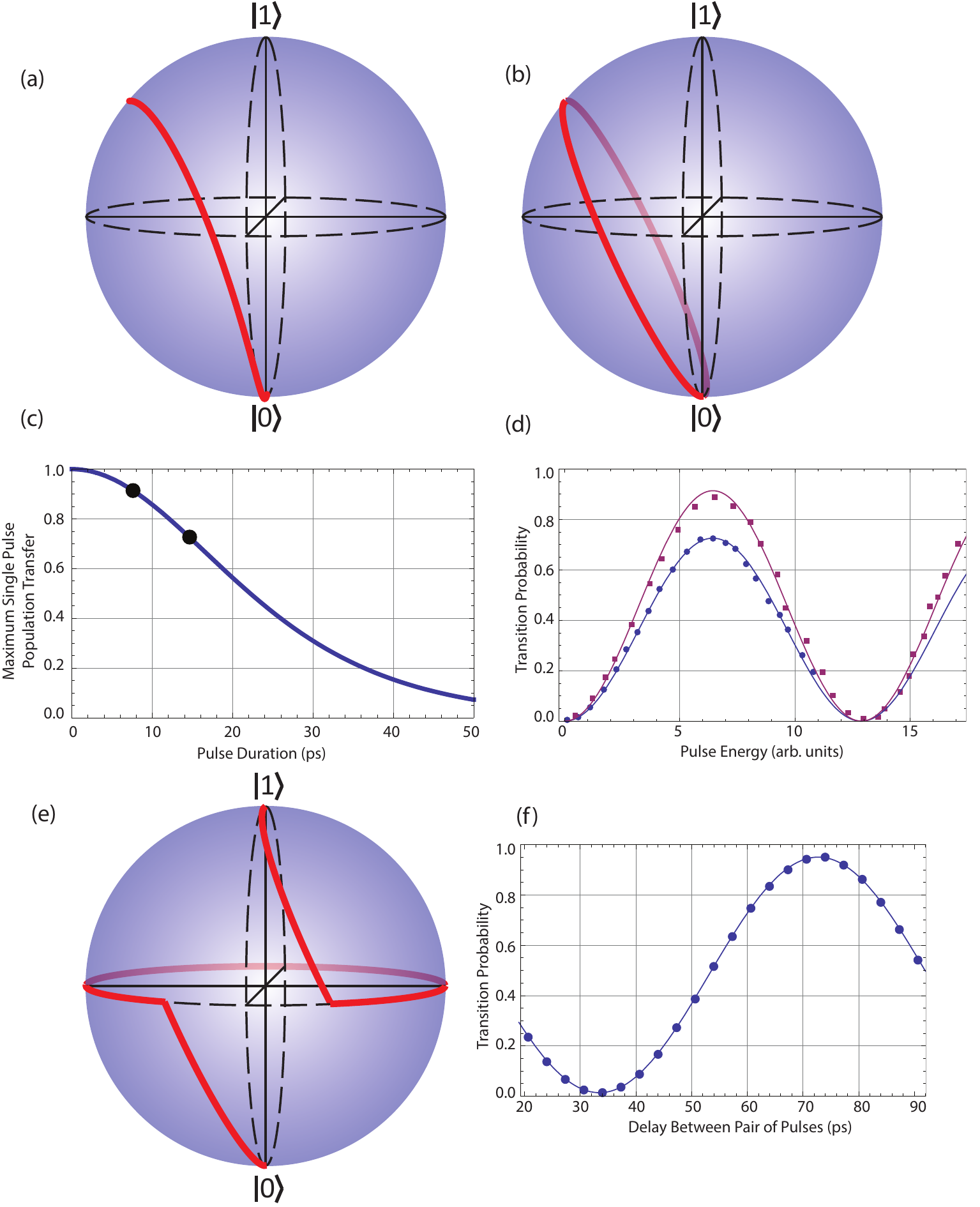}%
\caption{
(a) Bloch sphere position as a function of pulse energy, following the Rosen-Zener solution in equations \ref{eq:RosenZenerA}-\ref{eq:RosenZenerXi}.
(b) The final position reached by the twisted path in (a) can be represented by a single effective rotation axis and angle, as in equation \ref{eq:RosenZenerURotation}. The angle of rotation is given by $\varphi$; the axis is determined by $\theta$ and $\tp$.
(c) Theoretical maximum population transfer in $\yb$ for a single pulse as a function of pulse duration, based on equation \ref{eq:P01SinglePulse}. The black dots indicate the points corresponding to the data in (d).
(d) Experimental data showing the behavior described theoretically in (a)-(c). Ion state is measured as a function of incident pulse energy. The transfer probability reaches a maximum given by equation \ref{eq:P01SinglePulse}. The two different datasets correspond to two different lasers with different pulse durations. The fit to the data show that those pulse durations are 14.7 ps (circles) and 7.6 ps (squares). These points are indicated on the plot in (c). 
(e) Two identical pulses separated by an appropriate delay can fully transfer the population.  Each pulse has sufficient energy to rotate the state to the equator of the Bloch sphere.  The appropriate delay is approximately the qubit cycle time $2\pi/\oq$.  It is slightly smaller due to the off axis rotation caused by the Rosen-Zener dynamics. 
(f) Data showing the effect in (e). As the delay between the pulses is scanned, the transition probability goes from 0 to 1. The maximum is less than one due to detection errors. }
\label{fig:FastSpinFlips}%
\end{figure}

The Hamiltonian of Eq. \ref{eq:Hamiltonian} and \ref{eq:OnePulseRabiFreq} for the hyperbolic secant Rabi frequency envelope in time was solved exactly by Rosen and Zener\cite{RosenZener}. For the purposes of this analysis, we are not interested in the dynamics during the pulse, but only the resultant state after the pulse. The evolution operator for a pulse followed by free evolution for a time $T$ is given by\cite{Vitanov, Robiscoe}:
\begin{align}
U &= \left( \begin{array}{ccc}
A & iB^* \\
iB & A^* \end{array} \right) \label{eq:RosenZenerU}
\end{align}
where $A$ and $B$ are given by:
\begin{align}
A &= \frac{\Gamma^2\left(\xi\right)e^{i\oq T/2}}{\Gamma\left(\xi-\frac{\theta}{2\pi}\right)\Gamma\left(\xi+\frac{\theta}{2\pi}\right)}\label{eq:RosenZenerA} \\
B &= -\sin\left(\frac{\theta}{2}\right)\sech\left(\frac{\oq\tp}{2}\right)e^{-i\oq T/2}\label{eq:RosenZenerB} \\
\xi &= \frac{1}{2}+i\frac{\oq\tp}{2\pi}\label{eq:RosenZenerXi}
\end{align}
where $\Gamma(\xi)$ is the Gamma function. For a fixed value of $\theta$, this evolution operator can be written as a pure rotation operator:
\begin{equation}
\tilde{U} = e^{i \varphi\hat{n}\cdot\vec{\sigma}/2}
\label{eq:RosenZenerURotation}
\end{equation}
where the rotation axis $\hat{n}$ and rotation angle $\varphi$ are given by:
\begin{align}
  \cos\left(\frac{\varphi}{2}\right) &= \text{Re}\left(A \right) \label{eq:varphi} \\
  	n_z \sin\left(\frac{\varphi}{2}\right) &= \text{Im}\left(A\right) \label{eq:nz} \\
		(n_x+in_y) \sin\left(\frac{\varphi}{2}\right) &= B \label{eq:nxiny}
\end{align}
The equivalent pure Bloch sphere rotation is shown in figure \ref{fig:FastSpinFlips}(b). Equation \ref{eq:RosenZenerURotation} allows the evolution operator to quickly be extended to $N$ pulses equally spaced by a time $T$:
\begin{equation}
U_N = e^{i N\varphi\hat{n}\cdot\vec{\sigma}/2}
\label{eq:UN}
\end{equation}
If the ion is initialized to the state $\ket{0}$, then the transition probability after $N$ pulses is given by:
\begin{align}
\nonumber P_{0\rightarrow 1}&= \left|i\sin\left(\frac{N\varphi}{2}\right)\left(n_x+in_y\right)\right|^2 \\
&= \left(\frac{|B|^2}{\sin^2\left(\frac{\varphi}{2}\right)}\right)\sin^2\left(\frac{N\varphi}{2}\right) \label{eq:TimeDomRabi}
\end{align}

To understand the behavior described by the above equations, first consider the limit of an infinitesimally short pulse: $\tp = 0$. In that case, equation \ref{eq:RosenZenerA} and \ref{eq:RosenZenerB} become:
\begin{align}
A &= \cos\left(\frac{\theta}{2}\right)e^{i\oq T/2} \\
B &= -\sin\left(\frac{\theta}{2}\right)e^{i\oq T/2}
\end{align}
If the time between pulses satisfies the condition:
\begin{equation}
\oq T = 2\pi n\text{, }n\in\mathbb{Z}\label{eq:TimeDomCond}
\end{equation}
then equations \ref{eq:varphi}, \ref{eq:nz}, and \ref{eq:nxiny} show that $\varphi = \theta$, $n_z=n_y=0$, and $n_x=1$. In this case, the action of each pulse is rotation about the $x$-axis, by an angle equal to the pulse area. Equation \ref{eq:TimeDomRabi} then becomes:
\begin{equation}
 P_{0\rightarrow 1} = \sin^2\left(\frac{N\theta}{2}\right)
\label{eq:TimeDomRabiDeltaFunction}
\end{equation}
This equation shows that the behavior is discretized Rabi flopping.

Now consider non-zero pulse duration. Equation \ref{eq:TimeDomRabi} shows that for $N=1$, the transition probability reduces to:
\begin{equation}
P_{0\rightarrow 1} = |B|^2 = \sin^2(\theta/2)\sech^2(\oq\tp/2)
\label{eq:P01SinglePulse}
\end{equation}
Therefore for a single pulse, the maximum population transferred is $\sech^2(\oq\tp/2)$. This quantity is always less than one, meaning a single pulse cannot fully flip the spin of the qubit. However, for two pulses, equation \ref{eq:TimeDomRabi} can be made equal to 1, for particular values of the delay time $T$. If $T\ll 1/\oq$, then the correct delay condition will be a small correction to equation \ref{eq:TimeDomCond}.

This can be understood by examining the qubit evolution on the Bloch sphere. The Bloch sphere path for the Rosen-Zener solution is shown as a function of $\theta$ in figure \ref{fig:FastSpinFlips}(a). Note that the path is twisted -- for small values of $\theta$, the rotation axis is nearly purely about the $x$-axis. As $\theta$ increases, the amount of $z$ rotation also increases. If $\theta$ is fixed, the final state can be connected to the initial state by a pure rotation, which is shown in figure \ref{fig:FastSpinFlips}(b).  For non-zero pulse duration, the rotation axis is never purely in the x-y plane, meaning the north pole of the Bloch sphere is never reached. However, two pulses can fully flip the spin, so long as one pulse can reach the equator, as shown in figure \ref{fig:FastSpinFlips}(e). For $\yb$, the condition for two pulses to be able to fully transfer population is $\tp < 22$ ps.

These results show that two fast pulses can be used to rotate the state of a qubit extremely rapidly, in less than one qubit period. Moreover, these same pulses can be used to rotate the phase of a qubit (i.e., z-rotations on the Bloch sphere). To see this, again consider a pair of pulses as above. However, instead of choosing a delay such that equation \ref{eq:TimeDomRabi} equals 1, a delay is chosen such that it equals 0; i.e. $\varphi = \pi$. In that case, the evolution operator causes a phase shift of the qubit, controllable via the power of the pulses.

Figure \ref{fig:FastSpinFlips}(d) show experimental results for a single pulse. The data sets shown correspond to two different lasers with different pulse durations. The circles shows a maximum brightness of $72\%$, corresponding to a pulse duration of $\tp = 14.8\:\text{ps}$ in equation \ref{eq:P01SinglePulse}. The squares shows a maximum of $91\%$, corresponding to $\tp = 7.6\:\text{ps}$. These numbers are consistent with independent measurements of the pulse duration.

Figure \ref{fig:FastSpinFlips}(f) shows the results of scanning the delay between two pulses. The two pulses were created by splitting a single pulse from the laser, and directing the two halves of the pulse onto the ion from opposite directions, as described in \cite{Campbell}. (Note that while the pulses are directed onto the ion from opposite directions, there is no coupling to the ion's motion -- the pulses are not overlapped in time. There is therefore no possibility of momentum transfer.) The maximum occurs at a delay of $72\:\text{ps}$, slightly less than one qubit period. The maximum is less than one due to detection errors.

\begin{figure}%
\includegraphics[width=\columnwidth]{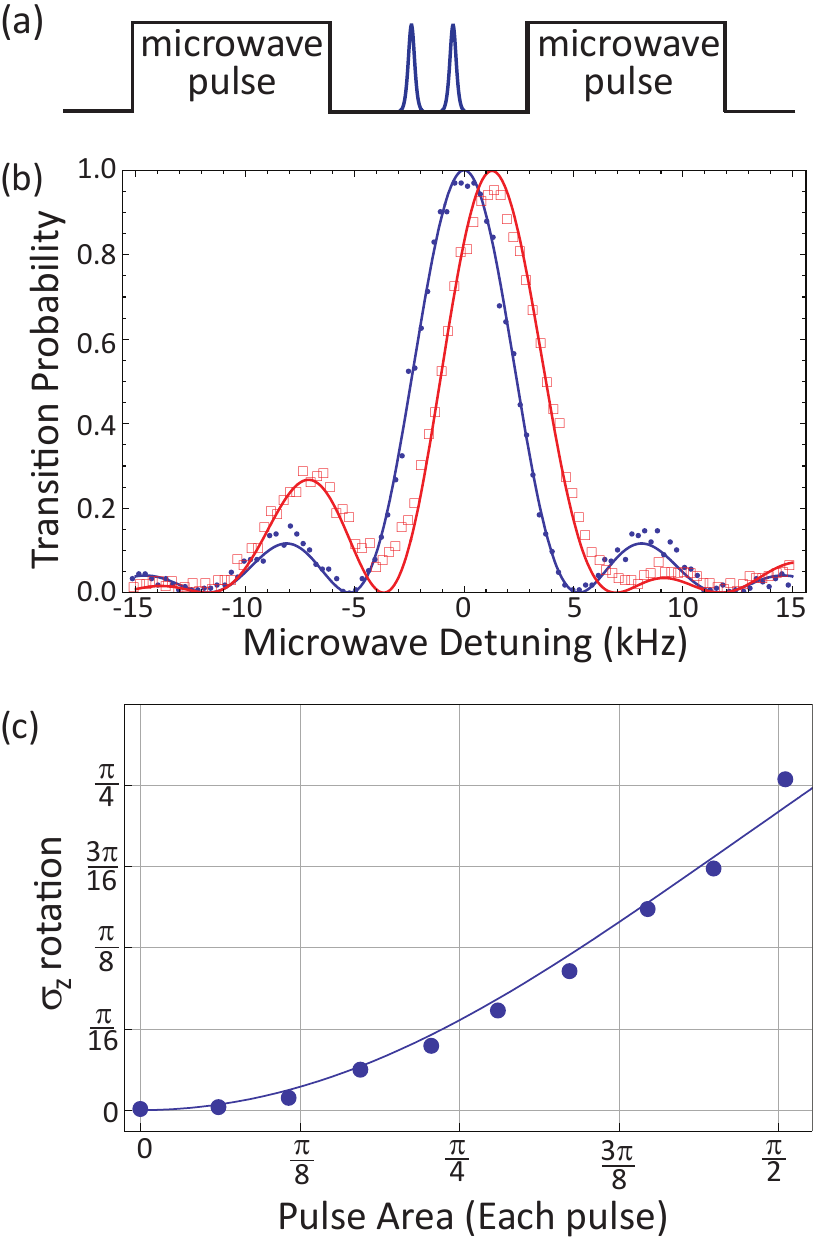}%
\caption{Data showing fast phase rotation caused by pair of pulses. (a) Ramsey sequence: the frequency of two microwave $\pi/2$ pulses is scanned. In between the microwaves, two fast laser pulses with delay set to cancel $x$ rotation are inserted. Fringe shift is then measured as a function of pulse area. (b) Data showing fringe shift. Circles: No laser pulses, Squares: Laser pulses of pulse area equal to $1.25$, showing phase rotation angle of $0.49$. (c) measured z-rotation angle as a function of pulse area.}
\label{fig:SigmaZ}%
\end{figure}

To demonstrate pure phase rotation, the delay between the pulses was set such that there was no net population transfer (34 ps delay in figure \ref{fig:FastSpinFlips}(f)). This pulse pair was then put between two $\pi/2$ Ramsey zones, and the frequency of those Ramsey zones scanned for different laser intensities. The phase shift caused by the laser pulses manifests as a shift in the Ramsey fringes. The angle of $z$ rotation can then be calculated based on the shift. The amount of phase rotation is set by controlling the intensity of the two pulses. The results are shown in figure \ref{fig:SigmaZ}. The fit curve in (c) is derived from the Rosen-Zener solution, equations \ref{eq:RosenZenerA}-\ref{eq:RosenZenerXi}. The only free parameter is the overall scaling, i.e. the correspondence between the measured pulse amplitude and the pulse area $\theta$ on the x-axis of the plot.

These results show that by controlling the intensity and delay between two fast pulses, any arbitrary Bloch sphere rotation can be achieved in tens of picoseconds.

\subsection{Weak Pulses}

In the above section, the pulse area was large, such that a single pulse had a significant effect on the qubit state. If instead the area per pulse is small ($\theta \ll 1$), then many pulses are required to coherently drive the qubit substantially. In this case, the analysis is better understood in the frequency domain. The Fourier transform of a train of equally spaced pulses with a fixed phase relationship is a frequency comb, with teeth spaced by the repetition frequency $\orep$. The width of an individual tooth in an $N$ pulse train scales like $\orep/N$. If the width of a tooth is small compared to the tooth spacing ($N\gg 1$), then the comb can be thought of as an ensemble of CW lasers.  All that remains is to ensure that the frequency comb spectrum includes optical beat notes that are resonant with the qubit splitting $\oq$.  

Note that since the qubit transitions are driven by a frequency difference between comb lines rather than by an absolute optical frequency, the carrier-envelope phase (CEP) is therefore irrelevant and does not need to be stabilized.
However, in order to coherently drive the qubit, it is important that the beat note at the qubit splitting be stable. In general, well-designed mode-locked lasers enjoy excellent passive stability of their repetition rate (comb tooth spacing) over the time scale of a coherent qubit operation (microseconds), so that individual operations are coherent.  Over longer times however, drifts in the repetition rate will spoil attempts to signal average or concatenate operations.  The fractional drift of the repetition rate, similar to the fractional linewidth and drift of a free-running CW laser, is typically in the range of $\sim 10^{-7}$ over minutes.  This drift can be eliminated by actively stabilizing the laser repetition rate, using a piezo mounted end mirror\cite{Hayes}. 

\subsubsection{Single Comb}
\begin{figure}%
\includegraphics[width=\columnwidth]{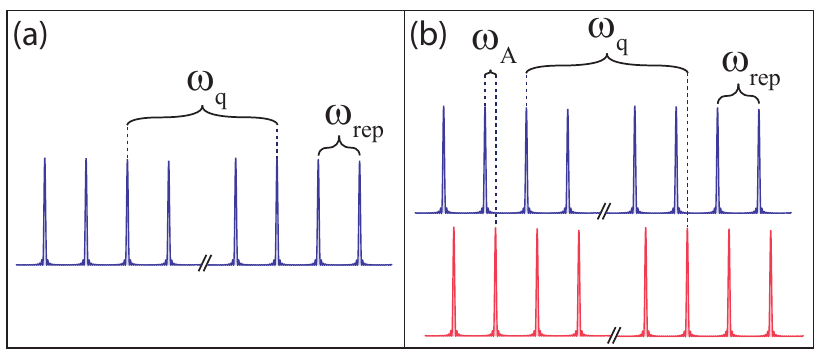}%
\caption{(a) One frequency comb can drive Raman transitions if pairs of comb lines are separated by the qubit frequency, leading to the condition in equation \ref{eq:FreqDomCond}.  (b)  Two frequency combs can drive Raman transitions together if a frequency offset $\oa$ between the combs causes lines from the separate beams to be spaced by the qubit frequency, leading to the condition in equation \ref{eq:FreqDomCond2}.}%
\label{fig:Combs}%
\end{figure}

A single comb of equally-spaced components can drive stimulated Raman transitions if the qubit splitting is an integer multiple of the comb teeth spacing, as shown in figure \ref{fig:Combs}(a):
\begin{equation}
\oq = n\orep\text{, }n\in\mathbb{Z}
\label{eq:FreqDomCond}
\end{equation}
This condition is equivalent to equation \ref{eq:TimeDomCond}. The Rabi frequency can be computed by summing the effect of all pairs of comb teeth separated by $\oq$\cite{Hayes}. For two CW phase-locked beams with single photon Rabi frequencies $g_1$ and $g_2$ (assumed to be real), the Raman Rabi frequency between qubit states is $\Omega = g_1g_2/2\Delta$.  For an optical frequency comb resulting from hyperbolic secant pulses, the $k$th comb tooth at frequency $k\orep$ from the optical carrier has single photon Rabi frequency
\begin{equation}
g_k=g_0 \sqrt{\frac{\orep \tp}{2}}\text{sech}(k\orep \tp) ,
\label{gk}
\end{equation}
where $g_0^2 = \sum_{k=-\infty}^{+\infty} g_k^2 =(\bar{I}/2I_{sat})\gamma^2$.  
The net two-photon Rabi frequency from the frequency comb is therefore
\begin{align}
\Omega &= \sum_{k=-\infty}^{+\infty} \frac{g_k g_{k+n}}{2\Delta} \\
&\approx \Omega_0 \sech\left(\frac{\oq\tp}{2}\right) ,
\label{eq:Omega1}
\end{align}
where $n$ is the number of comb teeth spanning the qubit splitting (Eq. \ref{eq:FreqDomCond}), $\Omega_0 = g_0^2/2\Delta$ and we assume the beatnotes at $\oq$ all add in phase since the pulse has negligible frequency chirp. The remaining hyperbolic secant factor is nearly unity when the individual pulse bandwidth $1/\tp$ is much greater than the qubit frequency splitting $\oq$.  

This expression can be connected to the time domain analysis above in a straightforward manner. In equation \ref{eq:TimeDomRabi}, the number of pulses $N$ can be replaced by time $t$ using the relation $N = 2\pi\orep t$. This shows that the Rabi frequency is related to the rotation angle $\varphi$ by:
\begin{equation}
\Omega = 2\pi\orep\varphi
\label{eq:OmegaVarPhi}
\end{equation}
Equation \ref{eq:TimeDomRabi} also shows that full contrast requires $\sin^2(\varphi/2) = B^2$, which is equivalent to the condition that the comb is driving the transition on resonance. This relation becomes:
\begin{align}
\sin^2\left(\frac{\varphi}{2}\right) &= \sin^2\left(\frac{\theta}{2}\right) \sech^2\left(\frac{\oq\tp}{2}\right)\label{eq:Omega3}\\
\Rightarrow \varphi &\approx \theta \sech\left(\frac{\oq\tp}{2}\right)\label{eq:Omega4}\\
\Rightarrow \Omega &= \Omega_0 \sech\left(\frac{\oq\tp}{2}\right)\label{eq:Omega5}
\end{align}
The second line follows from the small angle approximation, and the third line is the second multiplied by $2\pi\orep$. This shows that the constant $\Omega_0 = 2\pi\orep\theta$. From this it is clear that the approximation made in treating the pulse train as an ensemble of CW lasers is equivalent to the assumption that the effect of an individual pulse is small.

In addition to the resonant beat note at the qubit frequency, there will also be many beat notes at integer multiples of $\orep$ away from the qubit frequency from the multitude of comb teeth splittings. These other beat notes will lead to a shift in the qubit resonance and can be thought of as a higher order four photon AC Stark shift.  From Eq. \ref{eq:Omega1}, the strength of the beat note 
at $j\orep$ is characterized by its resonant Rabi frequency $\Omega_j \approx \Omega_0 \text{sech}  (j\orep\tau/2)$. 
The net four photon Stark shift is then a sum over all nonresonant beatnotes,
\begin{align}
\nonumber \delta_4 &=  -\sum_{\substack{j=-\infty\\j \neq n}}^{\infty}\frac{\Omega_j^2}{2(j\orep-\oq)} \\
                      									&= -\frac{\Omega_0^2}{2\orep}\sum_{\substack{j=-\infty\\j \neq 0}}^{\infty}
								                                                             \frac{\text{sech}^2[(j+n)\orep\tau/2]}{j} \\
			                                  & \approx 0.853\frac{\Omega_0^2\oq\tau}{\orep}
\label{eq:LightShiftComb}
\end{align}
The last expression is valid in the case where $\orep \tp \ll 1$ and to lowest order in $\oq\tp/2$.  For laser pulses of $\tp=10$ ps duration with a repetition rate $\orep/2\pi = 80$ MHz and net Rabi frequency $\Omega/2\pi = 1$ MHz, for the $\yb$ qubit we find a resultant $4$-photon Stark shift of $\delta_4/2\pi \approx +8.5$ kHz. It should be noted that equation \ref{eq:LightShiftComb} could also be derived from the time domain Rosen-Zener solution discussed in section \ref{sec:StrongPulseSpin}.

\subsubsection{Two Combs}
Equation \ref{eq:FreqDomCond} requires a laser with a repetition rate that is commensurate with the qubit splitting. 
However, this may be difficult to achieve in practice, and in any case it is undesireable for non-copropagating laser pulses -- such a laser cannot generate the spin-dependent forces discussed in section \ref{sec:SpinMotion}. Moreover, the repetition rate on many mode-locked lasers cannot be easily controlled to stabilize drifts. It is therefore convenient to generate two combs, with one frequency shifted relative to the other, typically via an AOM as shown in figure \ref{fig:Combs}(b).   In this way, Raman transitions are controlled through this frequency offset and this configuration allows atomic forces to be exerted in a given direction when tuned near motional sideband transitions (see section \ref{sec:SpinMotion}).  Finally, drifts in the repetition rate can be measured and fed forward onto a downstream modulator, in case the repetition rate of a laser is not accessible.  This feed-forward effectively eliminates drift in the relevant comb beatnote to drive qubit transitions by  \ref{eq:FreqDomCond2}, as the 
``breathing" of the comb teeth is compensated by the offset comb \cite{IslamThesis}.

\begin{figure}%
\includegraphics[width=\columnwidth]{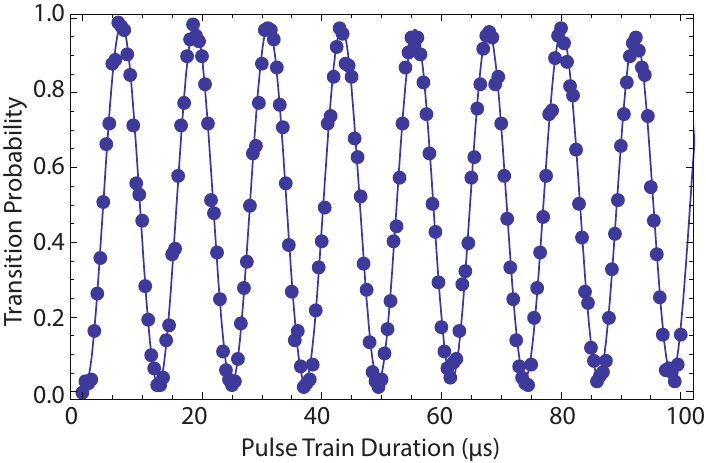}%
\caption{Rabi oscillations driven by a pair of copropagating combs with an AOM shift between them. In this data the laser repetition frequency is directly stabilized.}%
\label{fig:RabiFlopping}%
\end{figure}

Including an offset frequency $\oa$ between the two combs, the condition for driving transitions now becomes:
\begin{equation}
\oq = n\orep\pm\oa\text{, }n\in\mathbb{Z}
\label{eq:FreqDomCond2}
\end{equation}
In order to allow for the possibility of spin-dependent forces in a counter-propagating geometry, we exclude the offset frequency values 
$\oa  = k\orep$ or $(k+1/2)\orep$, $k\in\mathbb{Z}$.
Figure \ref{fig:RabiFlopping} shows Rabi flopping driven by two offset optical frequency combs, in a copropagating geometry where the repetition rate is directly stabilized.

The Rabi frequency for the case of two offset combs is exactly as written for the case of a single comb 
(Eq. \ref{eq:Omega1}), where this time $g_0^2 = (\bar{I}/2I_{sat})\gamma^2$ characterizes the intensity $\bar{I}$ 
of each of the two combs.
For the offset combs the four-photon AC Stark shift is modified from the asymmetry in the spectrum of two-photon beatnotes.  Once again summing over all nonresonant beatnotes, we find
\begin{widetext}
\begin{align}
\nonumber \delta_4 &=  -\sum_{\substack{j=-\infty\\j \neq n}}^{\infty}\frac{\Omega_j^2}{2(j\orep+\omega_A-\oq)} 
                                                -\sum_{j=-\infty}^{\infty}\frac{\Omega_j^2}{2(j\orep+\omega_A+\oq)}  \\
              &= -\frac{\Omega_0^2}{2\orep} \left[
							           \sum_{\substack{j=-\infty\\j \neq 0}}^{\infty} \frac{\text{sech}^2[(j+n)\orep\tau/2]}{j} 
												         - \sum_{j=-\infty}^{\infty} \frac{\text{sech}^2[(j-n)\orep\tau/2]}{j+2\sigma}   \right]  \\
			                                  & \frac{\Omega_0^2}{2\orep} \left[ 3.412\omega_q\tau 
									  + \text{sech}^2(\omega_q\tau/2)\left(\frac{1}{2\sigma}+\frac{1}{1+2\sigma}+\frac{1}{2\sigma-1}\right) \right]
\label{eq:LightShiftComb2}
\end{align}
\end{widetext}
where $\tilde{\omega}_A = \oa(\text{mod } \orep)$, and $\sigma \equiv \tilde{\omega}_A/\orep$ is the fractional number of comb teeth that the two combs are offet $(0<\sigma<1$ and $\sigma \ne 0.5)$, 
and again we assume $\orep \tp \ll 1$.  The extra terms in the Stark shift compared to the single comb case (Eq. \ref{eq:LightShiftComb}) account for the closer asymmetric beat notes.  Interestingly, the net 4-photon AC Stark shift can be nulled by choosing 
a particular offset frequency for a given pulse duration.  In the $\yb$ system for example, we find that a value of 
$\sigma \sim 0.35 (0.40)$ nulls the Stark shift for pulse duration $\tp\approx 5 (10)$ ps.  For infinitesimally short pulses 
$(\tp \rightarrow 0)$, the Stark shift vanishes at the value $\sigma = 1/\sqrt{12}$.

\section{\label{sec:SpinMotion}Entanglement of Spin and Motion}

\begin{figure}%
\includegraphics[width=\columnwidth]{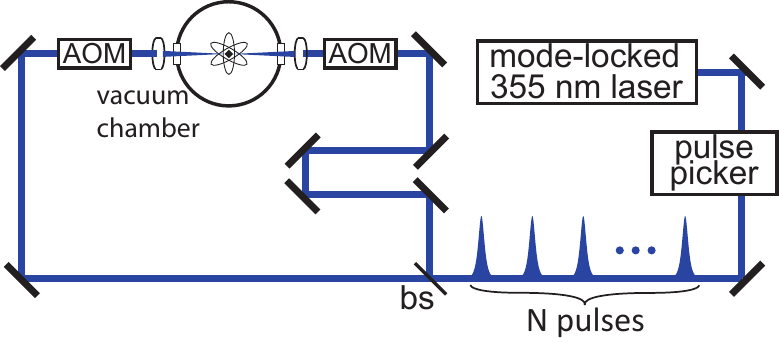}%
\caption{Experimental layout for counterpropagating geometry. The pulse train is split, and a frequency shift between the two arms is imparted by AOMs.}%
\label{fig:CounterProp}%
\end{figure}
The above section treated spin flips from copropagating pulses. Consider now a pair of counterpropagating pulse trains, as in figure \ref{fig:CounterProp}. The pulses are timed such that they arrive at the ion simultaneously, and the entire train has effective pulse area of order $\pi$. The frequency space picture is the same as in figure \ref{fig:Combs}(b) -- the two combs have a relative frequency shift, such that there exist pairs of comb lines that match the qubit splitting. However, absorption from one comb and emission into the other now leads to momentum transfer. Moreover, the direction of the momentum transfer is spin-dependent, leading to a spin-motion coupling. The form taken by that coupling will differ based on the duration of the pulse train. If the pulse train is much faster than the trap period, the result will be a spin-dependent kick: $\ket{0}$ and $\ket{1}$ will receive momentum kicks in opposite directions. If the pulse train is much slower than the trap period on the other hand, motional sidebands will be resolved.  In the Lamb-Dicke limit where the ion motion is confined much smaller than the optical wavelength, the motion will not be changed when on resonance, while a phonon will be added or subtracted when the beat note between the combs is detuned by the trap frequency.

To understand this process, first consider the effect of a single pair of pulses that arrive simultaneously on the ion from opposite directions. If the two pulses have orthogonal linear polarizations which are mutually orthogonal to the quantization axis ($\lpl$), then transitions can only be driven via the polarization gradient created by the two pulses. The Rabi frequency then acquires a sinusoidal position dependence. Under the instantaneous pulse approximation ($\tp=0$), the Hamiltonian for the ion-pulse interaction becomes:
\begin{equation}
\mathit{H} = -\frac{\theta}{2}\delta(t-t_0)\sin\left[\Delta k\hat{x} +\phi(t_0)\right]\sx
\label{eq:HamiltonianMotion}
\end{equation}
where $\theta$ is again the total pulse area, $t_0$ is the arrival time of the pulse pair, $\Delta k$ is the difference in wavevectors, $\hat{x}$ is the position operator for the ion, and $\phi(t_0)$ is the phase difference between the pulses. The time dependence of this phase difference comes from the AOM frequency shift:
\begin{equation}
\phi(t) = \oa t +\phi_0
\label{eq:Phitime}
\end{equation}
where $\phi_0$ is assumed to be constant over the course of one experiment. Equation \ref{eq:HamiltonianMotion} can be directly integrated to obtain the evolution operator for a single pulse arriving at time $t_0$:
\begin{align}
\nonumber U_{t_0} &= \exp\left(-i\int\mathit{H}(t)dt\right)\\
&=e^{i\frac{\theta}{2}\sin\left(\Delta k\hat{x} + \phi(t_0)\right)\sx} \label{eq:U_mot_pulse}\\
&=\sum_{n=-\infty}^{\infty} e^{in\phi(t_0)}J_n(\theta)D[in\eta]\sx^n\label{eq:UMotBessel}
\end{align}
where $J_n$ is the Bessel function of order $n$, $D$ is the coherent state displacement operator \cite{Glauber}, and $\eta$ is the Lamb-Dicke parameter.

\begin{figure}%
\includegraphics[width=\columnwidth]{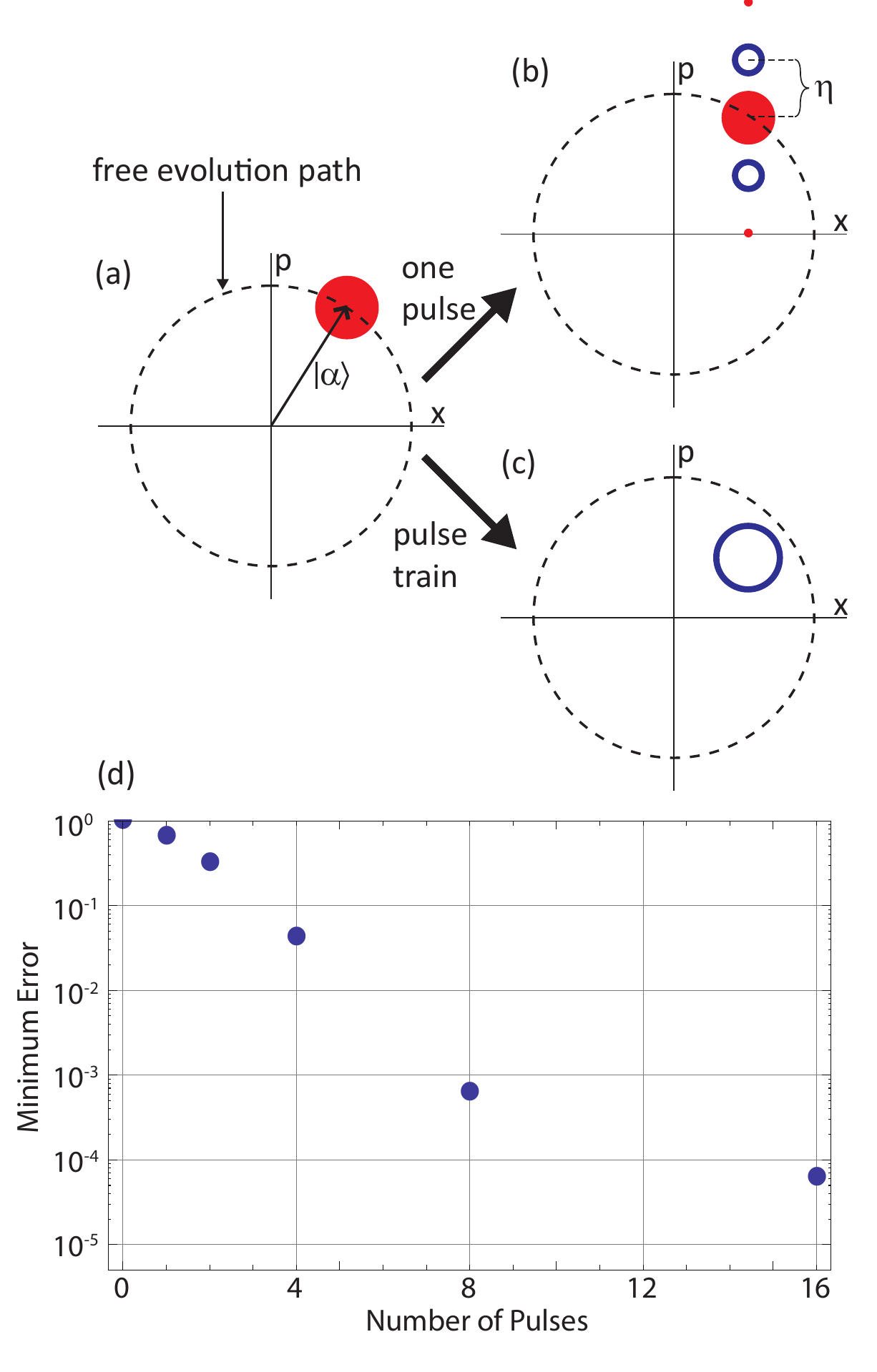}%
\caption{Phase space diagrams of pulse action.  Red closed circles indicate $\ket{0}$, while blue open circles indicate $\ket{1}$.  The size of the circle indicates the population in that state.  (a) Phase space diagram of an ion initially in the state $\ket{0}\ket{\alpha}$.  (b) Upon the arrival of a pulse pair, the ion is diffracted into a superposition of states as in equation \ref{eq:UMotBessel}.  (c) After $N$ pulse pairs satisfying equation \ref{eq:qpm}, population coherently accumulates in the state $\ket{1}\ket{\alpha-i\eta}$ and no other state, as in equation \ref{eq:StrongOpFinalPi}.  Similarly, population initially in $\ket{1}\ket{\alpha}$ coherently accumulates in $\ket{0}\ket{\alpha+i\eta}$. (d) Theoretical error (1-fidelity) of (c) as a function of $N$.  The convergence is very fast -- 4 pulses is $96\%$, 8 pulses $99.9\%$, and 16 pulses $99.99\%$.}%
\label{fig:Kicks}%
\end{figure}

Equation \ref{eq:UMotBessel} consists of operators of the form $D[in\eta]\sx^n$, which impart $n$ momentum kicks of size $\eta$ together with $n$ spin flips.  Physically, this corresponds to the process of absorbing a photon from one beam, emitting a photon into the other beam, repeated $n$ times. Each process of absorption followed by emission changes the momentum by $\eta$. The amplitude for the $n^\text{th}$ such process is given by the Bessel function $J_n(\theta)$, together with a phase factor.  The net action of this operator on a spin state $\ket{0}$ and coherent motional state $\ket{\alpha}$ is therefore to create a superposition of states of different size kicks, with alternating spin states.  This is shown graphically in figure \ref{fig:Kicks}(b).

This behavior can be understood as the scattering of the atomic wavepacket off of the standing wave of light, known as Kapitza-Dirac scattering\cite{KapitzaDirac,Sapiro,Gould}.  It has been directly observed in atomic beams\cite{Sapiro,Gould}.  It is also similar to the behavior observed in $\delta$-kicked rotor experiments\cite{Currivan}, although complicated by the presence of the spin operator. 

The evolution operator $O_N$ for a train of $N$ pulses will consist of a sequence of operators of the form \ref{eq:UMotBessel}, separated by free evolution:
\begin{equation}
O_N = U_{t_N} \ldots \ufe(t_3-t_2)U_{t_2}\ufe(t_2)U_{t_1}
\label{eq:NPulseOp}
\end{equation}
where $t_n$ is the arrival time of the $\nth$ pulse, and $\ufe(T)$ is the free evolution operator for time $T$, given by:
\begin{equation}
\ufe(T) = e^{-i\otrap T a^\dagger a}e^{-i\oq T\sz/2}
\label{eq:ufe}
\end{equation}

Let the total pulse train area be given by $\Theta$, so that a single pulse area is $\theta = \Theta/N$.  Assume that $N$ is sufficiently large such that the single pulse evolution operator in equation \ref{eq:UMotBessel} can be approximated to first order in $1/N$:
\begin{equation}
U_{t_k} \approx 1 + \frac{i\Theta}{2N}\left(e^{i\phi(t_k)} D\left[i\eta\right]+e^{-i\phi(t_k)} D\left[-i\eta\right]\right)\sx
\label{eq:UMotBesselApprox}
\end{equation}
Without loss of generality, assume $t_1 = 0$. Transforming to the interaction picture, $U_{t_k}$ becomes:
\begin{align}
\nonumber V_{t_k} &= \ufe^\dagger(t_k) U_{t_k} \ufe(t_k) \\
\nonumber &= 1 + \frac{i\Theta}{2N}\bigg\{e^{i\phi_0} D\left[i\eta e^{i\otrap t_k}\right]\times\\
&\left(e^{iq_+ t_k}\splus+e^{iq_- t_k}\sminus\right) +\text{H.c.}\bigg\} \label{eq:TransformedOp} \\
q_\pm &= \oq\pm\oa
\end{align}
Under this transformation, the interaction picture pulse train operator becomes:
\begin{equation}
\tilde{O}_N = \prod_{k=N}^1 V_{t_k}
\label{eq:TransformedTrainOp}
\end{equation}
There will now be two different approximations made in the fast regime ($\otrap t_N\ll 1$) and the slow regime ($ \otrap t_N\gg 1$).

\subsection{Fast Regime}
In the fast regime, $\otrap\approx 0$ during the pulse train, so that the ion is effectively frozen in place.  Equation \ref{eq:TransformedOp} then becomes:
\begin{equation}
V_{t_k} = 1 + \frac{i\Theta}{2N}\left(e^{i\phi_0} D\left[i\eta\right]\left(e^{iq_+ t_k}\splus+e^{iq_- t_k}\sminus\right) + \text{H.c.}\right)
\label{eq:StrongV}
\end{equation}
Now consider the product in equation \ref{eq:TransformedOp}. Suppose $q_\pm t_k\in\mathbb{Z}$, for all pulses, while $q_\mp t_k\notin\mathbb{Z}$. In frequency space, this is equivalent to satisfying one of the resonance conditions in equation \ref{eq:FreqDomCond2}, but not the other. The $q_\pm$ terms in the product in equation \ref{eq:StrongV} will then coherently add, while the $q_\mp$ terms will not. As the number of pulses grows, the non-resonant terms are strongly suppressed.  In frequency space, the comb lines narrow with increasing $N$, resulting in decreased amplitude for non-resonant processes. For large numbers of pulses on resonance, equation \ref{eq:StrongV} becomes:
\begin{equation}
V_{t_k} = 1 + \frac{i\Theta}{2N}\left(e^{i\phi_0} D\left[i\eta\right]\spm + \text{H.c.}\right)
\label{eq:StrongVresonant}
\end{equation}
The pulse train operator is now a product of identical operators:
\begin{align}
\nonumber &\tilde{O}_N = \left(1 + \frac{i\Theta}{2N}\left(e^{i\phi_0} D\left[i\eta\right]\spm + \text{H.c.}\right)\right)^N \\
& \xrightarrow{N\rightarrow\infty} \exp\left(\frac{i\Theta}{2}\left(e^{i\phi_0} D\left[i\eta\right]\spm + \text{H.c.}\right)\right) \label{eq:ExpRule}\\
&=\cos\frac{\Theta}{2}+i\sin\frac{\Theta}{2}\left(e^{i\phi_0} D\left[i\eta\right]\spm+e^{-i\phi_0} D\left[-i\eta\right]\smp\right) \label{eq:StrongOpFinal}
\end{align}
For a total pulse area of $\Theta = \pi$, Equation \ref{eq:StrongOpFinal} becomes:
\begin{equation}
\tilde{O} = ie^{i\phi_0} D\left[i\eta\right]\spm+ie^{-i\phi_0} D\left[-i\eta\right]\smp
\label{eq:StrongOpFinalPi}
\end{equation}
This is a spin-dependent kick operator. This shows that if the following conditions are satisfied:
\begin{align}
\nonumber \frac{q_\pm t_k}{2\pi}&\in\mathbb{Z}\\
\frac{q_\mp t_k}{2\pi}&\notin\mathbb{Z}\label{eq:qpm}
\end{align}
then a pulse train will create a spin-dependent kick, with the direction of kick determined by the sign chosen. Note that this result does not depend on being in the Lamb-Dicke regime. If the pulses are equally spaced, then $t_k = 2\pi k/\orep$, and equation \ref{eq:qpm} is equivalent to equation \ref{eq:FreqDomCond2}.

While the above analysis shows convergence to a spin-dependent kick in the limit of infinite pulses, it does not show how fast that convergence happens. Numerical analysis shows that it is quite fast, with better than $99.9\%$ fidelity after only 8 pulses. Figure \ref{fig:Kicks}(d) shows a numerical simulation of the maximum achievable fidelity with $N$ pulses.

It is clear from the time domain analysis that these pulse do not have to be equally spaced. Indeed, numerical optimization shows that the best fidelity is achieved for unequally spaced pulses.  To understand this result, consider the lowest order terms in the product in equation \ref{eq:TransformedTrainOp}. To first order, the coefficient of the $D\left[i\eta\right]\splus+\text{H.c.}$ term will be:
\begin{equation}
\sum_{k=1}^N e^{i q_+ t_k}
\label{eq:qp}
\end{equation}
while the coefficient of the term $D\left[i\eta\right]\sminus+\text{H.c.}$ will be:
\begin{equation}
\sum_{k=1}^N e^{i q_- t_k}
\label{eq:qm}
\end{equation}
The first order resonance requirement is then that one of these equations be maximal, which is the requirement in equations \ref{eq:qpm}. The second order requirement is that the other equation be zero, meaning there is complete destructive interference for the opposite direction kick. This is a separate requirement imposed on the pulse arrival times. Indeed, there will also be higher order corrections, further suppressing unwanted terms.

\begin{figure}%
\includegraphics[width=\columnwidth]{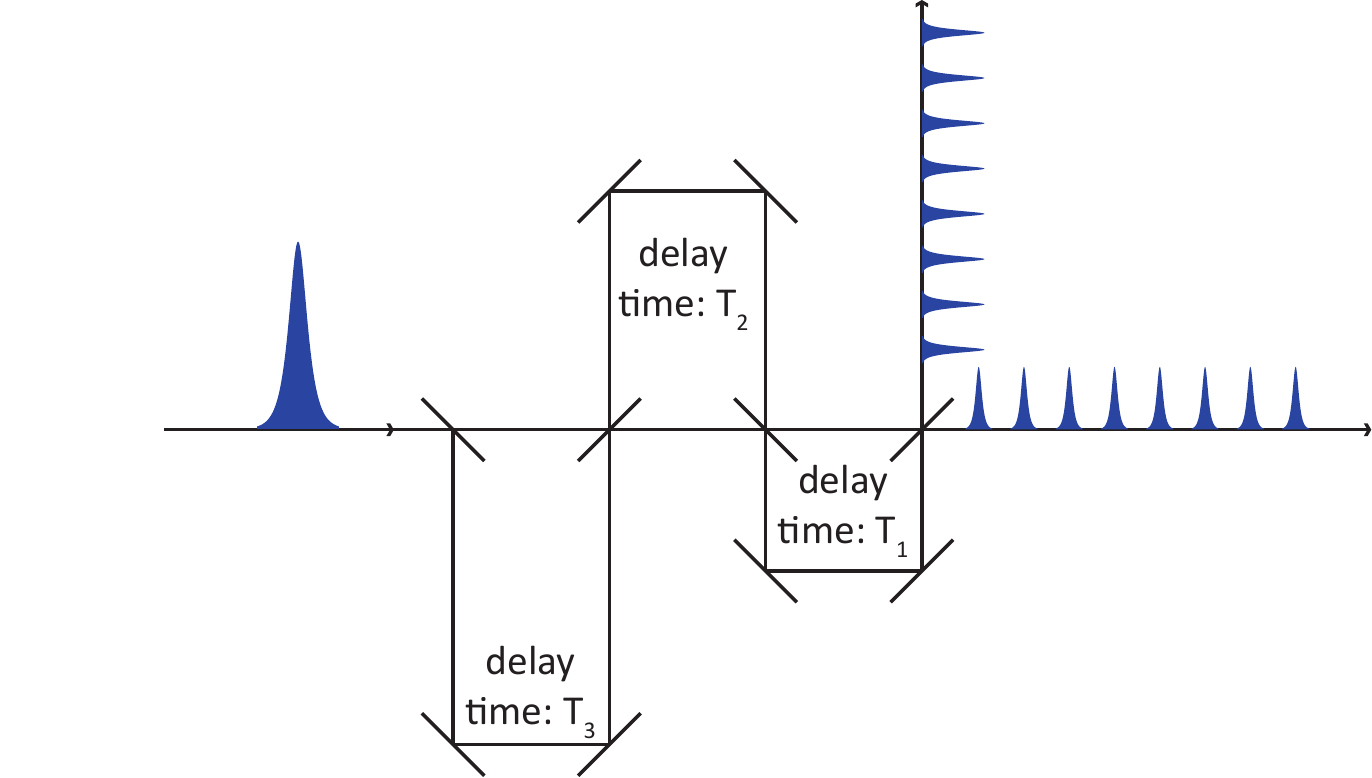}%
\caption{Optical layout for creating fast pulse train from a single pulse.}%
\label{fig:DelayLines}%
\end{figure}

In order for the approximation $\otrap \approx 0$ to be valid, the duration of the pulse train must be at least 2-3 orders of magnitude shorter than the trap period. A typical trap period if of order $1\:\mu\text{s}$, meaning the pulse train cannot be longer than a few nanoseconds. However, the repetition rate of pulses produced by the available lasers is only 80-120 MHz. At that rate, the ion would experience significant trap evolution even over the course of a small number of pulses. As an alternative, a single pulse from the laser followed by a sequence of delay lines can create a very fast pulse train, as shown in figure \ref{fig:DelayLines}. The limitation on the speed is then determined by the AOM frequency.

We demonstrated in \cite{Mizrahi} the creation of a spin-dependent kick of the form in equation \ref{eq:StrongOpFinalPi}. There, we showed that such kicks entangle the spin with the motion, while a second kick can disentangle the motion at integer multiples of the trap period.

Direct observation of the motional state of a trapped ion is extremely difficult, and motional information is typically determined by mapping to the spin\cite{Leibfried}.  Therefore, to detect that we created the operator in equation \ref{eq:StrongOpFinalPi}, it is necessary to infer the motional entanglement from its impact on the measured spin state.  To do this, we performed a Ramsey experiment using microwaves.  The experimental sequence was: (1) Initialize the spin state to $\ket{0}$, (2) Perform a $\pi/2$ rotation using near resonant microwaves, (3) Perform a spin-dependent kick using a single pulse through the interferometers, (4) Wait a time $T_\text{delay}$ (5) Perform a second spin-dependent kick, (6) Perform a second $\pi/2$ rotation, and (7) measure the state of the ion.  The frequency of the microwaves was then scanned. If the motion is disentangled from the spin, the result should be full contrast of the Ramsey fringe. On the other hand, if the spin and motion are entangled, then the trace over the motion will destroy the phase coherence. The result will be no Ramsey fringes. The motion should disentangle when $T_\text{delay}$ matches an integer multiple of the trap frequency.

\begin{figure}%
\includegraphics[width=\columnwidth]{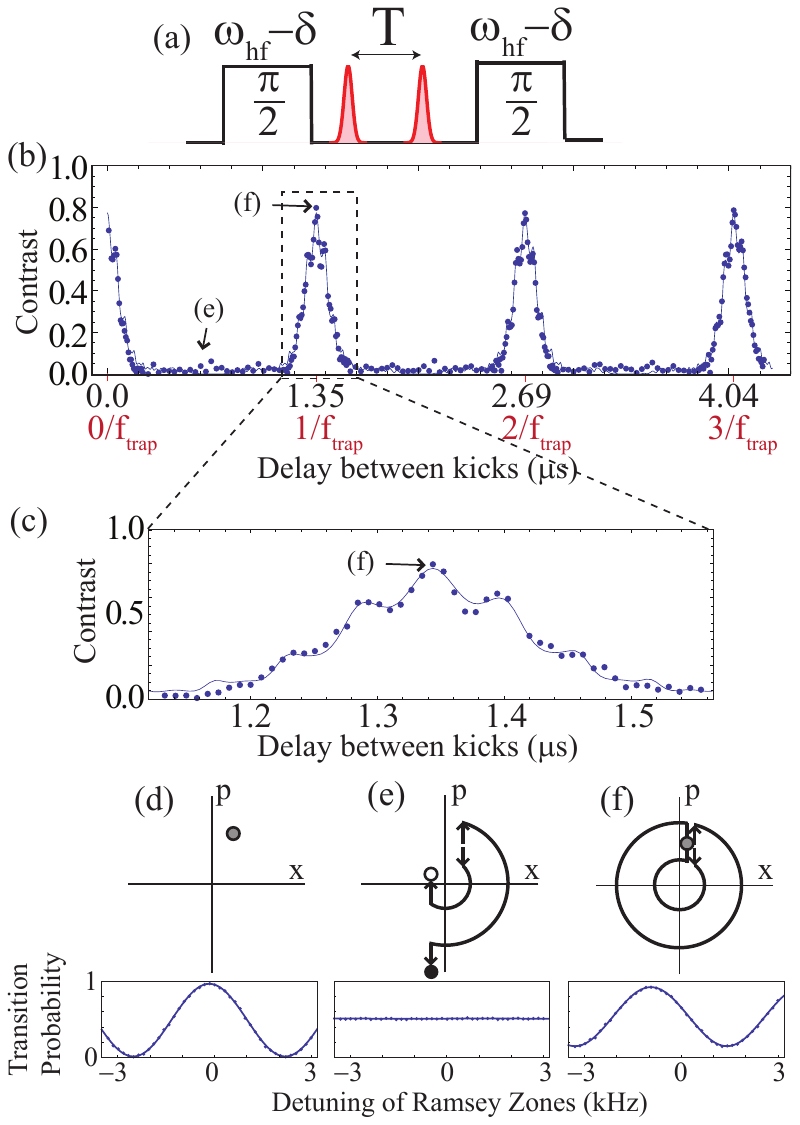}%
\caption{[Reproduced from \cite{Mizrahi}]. (a) Ramsey experiment to measure effect of spin-dependent kicks. Two spin-dependent kicks, separated by a time $T$ are placed between two microwave $\pi/2$ pulses. (b) Ramsey contrast as a function of delay between kicks. Clear revivals of contrast are seen at integer multiples of the trap period. (c) Close up of one revival peak. The small modulation present in the peak is due to uncompensated micromotion. The width of the peak is a function of the ion temperature and the micromotion amplitude. (d)-(f) phase space representation at various points on the plot in (a). Also shown are the Ramsey frequency scans at those points, showing the presence or lack of contrast.}%
\label{fig:ContrastRevival}%
\end{figure}

Figure \ref{fig:ContrastRevival} shows the results of this experiment. Plotted is the Ramsey contrast as a function of $T_\text{delay}$. The clear collapse and revival of contrast is a strong indicator that the pulses are indeed performing the spin-dependent kick in equation \ref{eq:StrongOpFinalPi}.  This sort of interferometry is similar to that discussed in \cite{Poyatos}.

\subsection{Slow Regime}
\begin{figure}%
\includegraphics[width=\columnwidth]{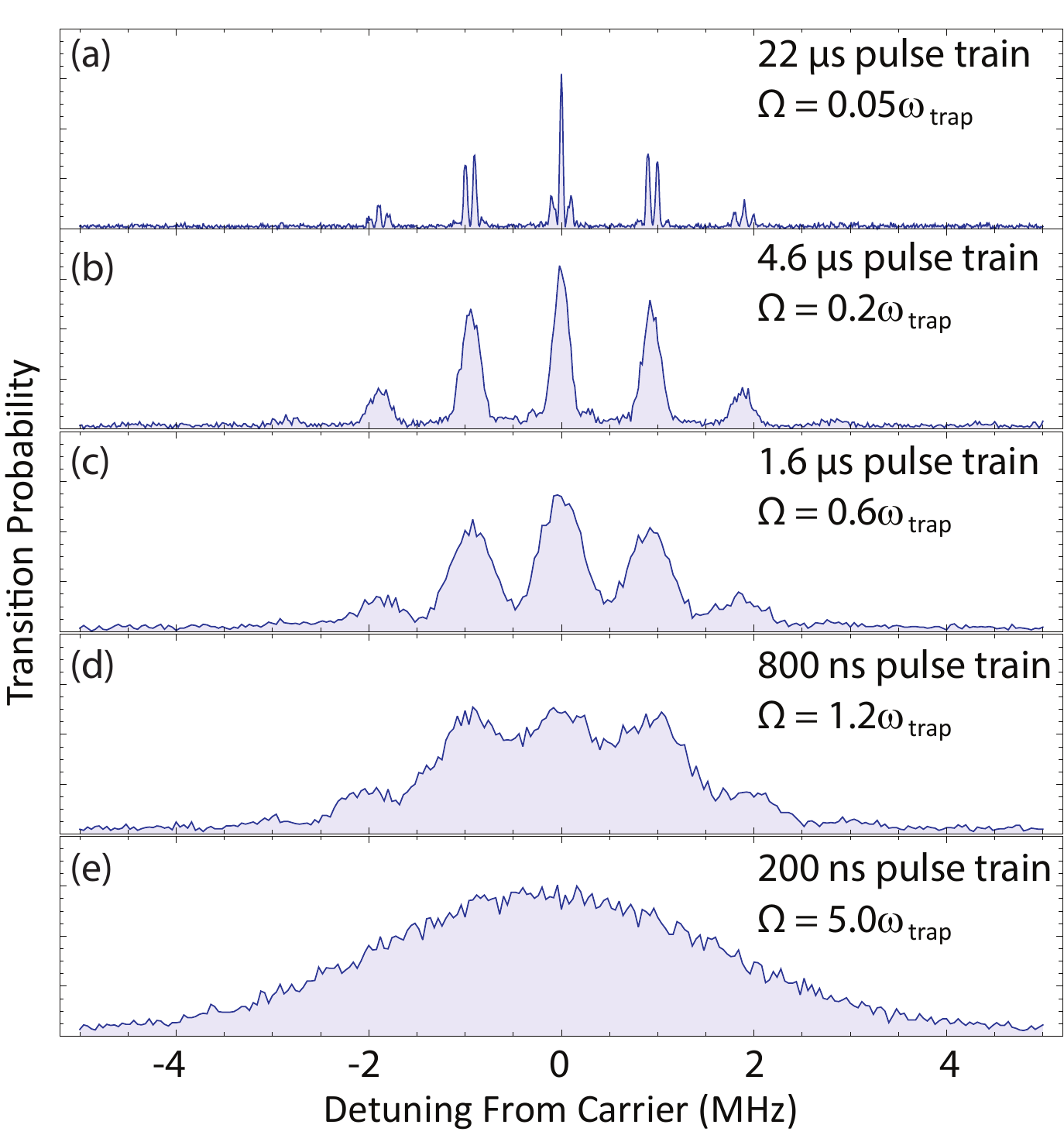}%
\caption{Data showing the crossover between the slow, resolved sideband regime and the fast, impulsive regime.  Each plot corresponds to scanning the frequency of an AOM in one of the arms of counterpropagating pulse trains.  In (a), $\Omega \ll \omega_t$, and sidebands transitions are clearly resolved.  As the pulse train power is turned up and the Rabi frequency increases, the lines begin to blur together.  In (e), no features are resolved at all, meaning all sidebands are being driven.}
\label{fig:TransitionSpectra}%
\end{figure}

In the slow regime, the pulse train is much longer than the trap cycle time: $t_N \gg 1/\otrap$. Now assume that the ion is in the Lamb-Dicke regime: $\eta\sqrt{\bar{n}+1} \ll 1$. In this regime, the following approximation can be made:
\begin{equation}
D\left[i\eta e^{i\otrap t_k}\right]\approx 1 + i\eta\left(e^{i\otrap t_k}a^\dagger+e^{-i\otrap t_k}a\right)
\label{eq:Dapprox}
\end{equation}
where $a$ and $a^\dagger$ are the harmonic oscillator annhilation and creation operators. Substituting this approximation into equation \ref{eq:TransformedOp} yields:
\begin{align}
\nonumber &V_{t_k}= 1 + \frac{i\Theta}{2N}\times \\
\nonumber &\bigg\{e^{i\phi_0}\left(1 + i\eta\left(e^{i\otrap t_k}a^\dagger+e^{-i\otrap t_k}a\right)\right)\times \\
&\left(e^{iq_+t_k}\splus+e^{iq_-t_k}\sminus\right) + \text{H.c.}\bigg\} \label{eq:VWeak}
\end{align}
There are now six phases to consider, associated with six different operators: $e^{iq_\pm t_k}$, $e^{i(q_\pm+\otrap)t_k}$, and $e^{i(q_\pm-\otrap)t_k}$.  The situation is then similar to the strong pulse regime: If one othese phases satisfies resonance (i.e. equal to 1 for all $t_k$) while the others do not, then the other terms will be negligible in the limit of large numbers of pulses.
For example, suppose that $(q_+ + \otrap)/2\pi\in \mathbb{Z}$, while none of the other phase terms satisfy this condition.  In that case, equation \ref{eq:VWeak} becomes:
\begin{equation}
V_{t_k} = 1 + \frac{i\Theta\eta}{2N}\left(ie^{i\phi_0}a^\dagger\splus - ie^{-i\phi_0}a\sminus\right)
\label{eq:VWeakResonant}
\end{equation}
As in the fast regime, the pulse train operator in equation \ref{eq:TransformedTrainOp} is now the product of identical operators, and converges to:
\begin{equation}
\tilde{O} = \cos\frac{\Theta\eta}{2}+i\sin\frac{\Theta\eta}{2}\left(ie^{i\phi_0}a^\dagger\splus - ie^{-i\phi_0}a\sminus\right)
\label{eq:WeakOpFinal}
\end{equation}
This is Rabi flopping on the blue sideband.  Similarly, the other resonance conditions correspond to red sideband and carrier operations. This behavior is shown in figure \ref{fig:TransitionSpectra}(a).

We previously reported in \cite{Hayes} on using pulse trains to do resolved sideband operations, as described above. There we demonstrated sideband cooling and two ion entanglement using the M\o{}lmer-S\o{}rensen technique\cite{Sorensen, Milburn, Haljan}.

Figure \ref{fig:TransitionSpectra} is experimental data showing the crossover between the slow and fast regimes. In this data, the transition probability was measured as a function of AOM detuning. In (a), sideband features are clearly resolved.  The peaks correspond to the carrier and sidebands at each of the three trap frequencies (1.0, 0.9, 0.1) MHz. These transitions follow from equation \ref{eq:VWeak}. As the power is increased and the pulse train duration decreased, the sidebands become less resolved, as the behavior crosses over from the slow regime to the fast regime. In (e), all of this structure has been washed out, and the motional transition is now described by impulsive kicks. From a sideband perspective, all sidebands are being driven simultaneously.

\section{\label{sec:Gates}Ultrafast Gates}

The goal of creating spin-dependent kicks of the form in equation \ref{eq:StrongOpFinalPi} is to execute a fast two ion entangling gate.  Such a gate would not be based on sidebands, and would therefore be fundamentally different from previously implemented two ion gates.  Because it does not depend on addressing sidebands, such a gate will be temperature insensitive, and would not require the ion to be cooled to the motional ground state or even cooled to the Lamb-Dicke regime.  Additionally, the Raman lasers generating the spin-dependent kick can be focused down to address just two adjacent ions in a long chain. If the gate is sufficiently fast, the other ions do not participate in the interaction. In principle, this allows this type of gate to be highly scalable. There have been theoretical proposals for such a gate in \cite{GarciaRipoll} and in \cite{Duan}.  Both schemes rely on using a sequences of spin-dependent kicks, timed such that the collective motion returns to its original state at the end of the process. This leaves a spin-dependent phase.

To understand the origin of this spin-dependent phase, consider a simple sequence of three spin-dependent kicks applied to two ions:
\begin{enumerate}
	\item $t=0$: momentum kick of size $+ \Delta k$
	\item $t=T$: momentum kick of size $-2 \Delta k$
	\item $t=2T$: momentum kick of size $+ \Delta k$
\end{enumerate}
This is a simplified version of the scheme proposed by Duan \cite{Duan}. Suppose that the total length of the kicking sequence is much faster than the trap period: $\otrap T \ll 1$. In that case, trap evolution during the kicks can be ignored, and the ions behave as free particles. The first kick imparts a momentum to each ion of $\Delta k$. The ions then move at a constant velocity away from equilibrium, until the second kick reverses the direction. The third kick then stops the motion of the ions at (nearly) the original position. 

\begin{figure}%
\includegraphics[width=\columnwidth]{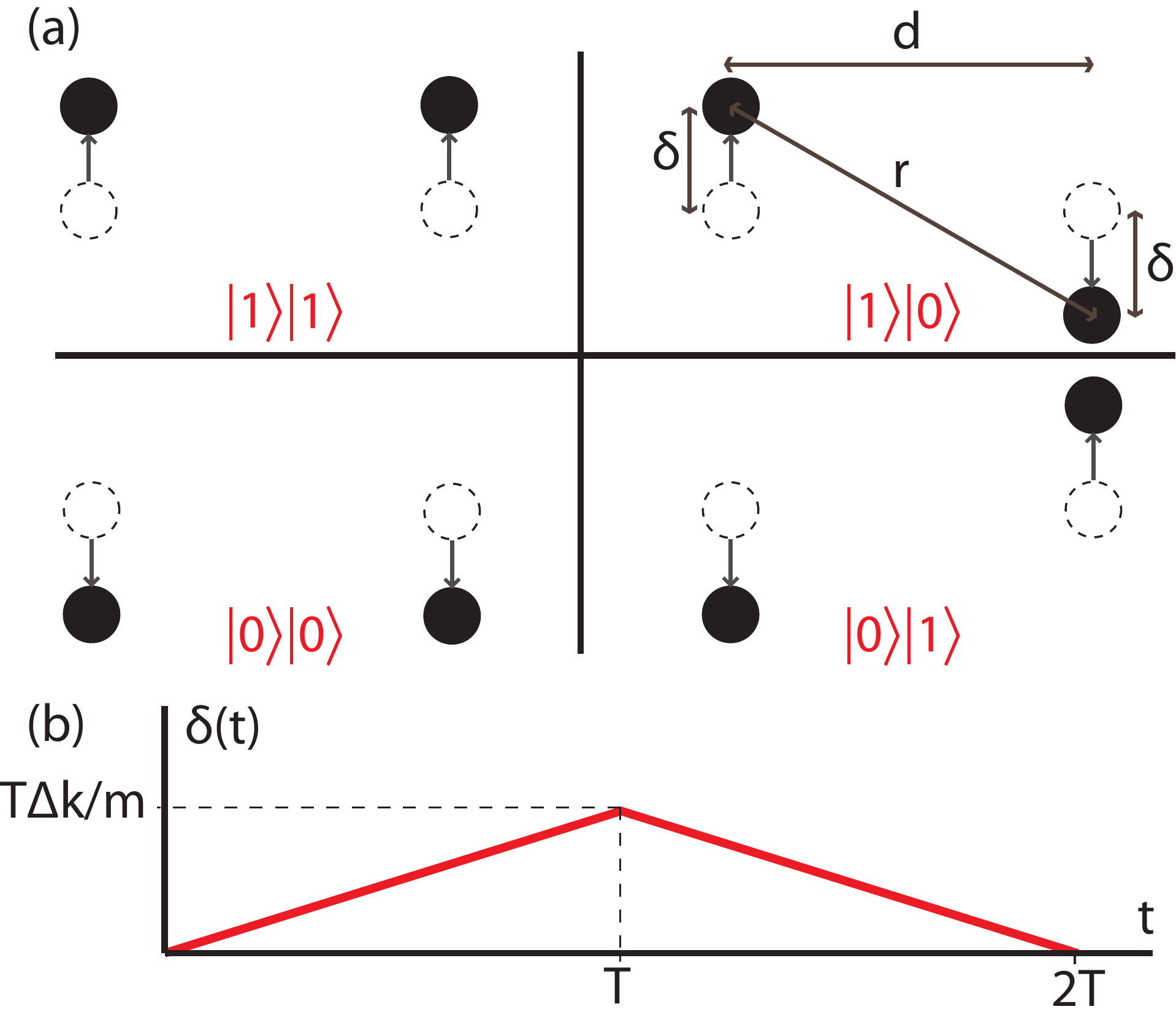}%
\caption{(a) The ground state of the motion is excited into four different possible configurations depending on the two ion spin state.  The dashed circles shows the original, equilibrium position of the ions.  The arrow and solid circles show the path followed after the first kick.  (b) In the limit where the kicks are much faster than the trap period, the trap evolution during the kicking sequence is negligible, and the ions can be considered as free particles.  The displacement $\delta$ of each ion from equilibrium as a function of time is shown.}%
\label{fig:CoulombPicture}%
\end{figure}

For two ions, there are four different possible spin states. Each will have a different motional excitations in response to these kicks, as shown in figure \ref{fig:CoulombPicture}(a).
If the ion spin state is $\ket{0}\ket{0}$ or $\ket{1}\ket{1}$, the two ion energy from the Coulomb interaction does not change during the sequence.  However, for $\ket{0}\ket{1}$ and $\ket{1}\ket{0}$, the energy changes as the ions get further apart and then closer together.  The time-dependent energy difference between these two configuration is:
\begin{equation}
\Delta E(t) = \frac{e^2}{d}-\frac{e^2}{\sqrt{d^2+\delta(t)^2}}\approx \frac{2e^2\delta(t)^2}{d^3}\label{eq:DeltaE}
\end{equation}
where $e$ is the electron charge, $d$ is the distance between the ions in equilibrium, and $\delta(t)$ is the displacement of each ion from equilibrium as a function of time (see figure \ref{fig:CoulombPicture}(b)).  The acquired phase difference from this process is given by:
\begin{equation}
\Delta\phi = \int_0^{2T} \Delta E(t) dt = \frac{4e^2\Delta k^2T^3}{3d^3m^2} \label{eq:DeltaPhi}
\end{equation}
We see then that the motional state (nearly) returns to its original state at the end of the process, while $\ket{0}\ket{1}$ and $\ket{1}\ket{0}$ acquire a phase relative to $\ket{0}\ket{0}$ and $\ket{1}\ket{1}$.  This is thus a phase gate. Note that the motion is entirely driven. Equation \ref{eq:DeltaPhi} is valid only because the ions are effectively free particles. The natural harmonic motion in the trap does not lead to phase accumulation. 

The fidelity of the phase gate described above is limited by free evolution in the trap. Because the gate is not truly instantaneous, there will be a small amount of residual entanglement with the motion at the end of the process. This infidelity can be eliminated by more complex kicking sequences, described below.

\begin{figure}%
\includegraphics[width=\columnwidth]{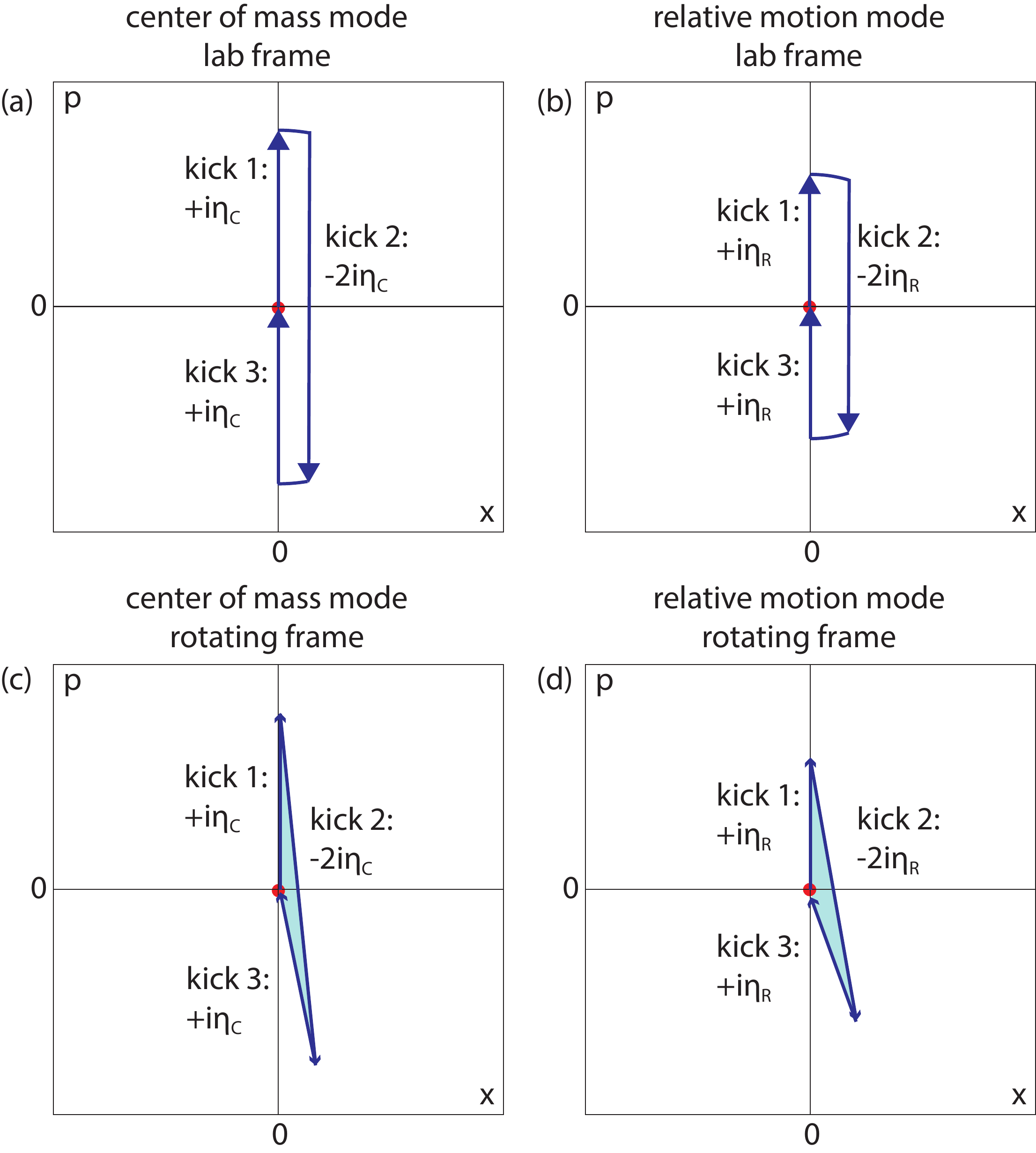}%
\caption{Phase space picture of the kick sequence described in the text. (a) and (b) are shown in the non-rotating frame, where free evolution follows circles in phase space. (c) and (d) are in the rotating frame. The phase difference is given by twice the difference in the enclosed area.}%
\label{fig:PhaseSpacePicture}%
\end{figure}

Alternatively, this process can be viewed as exciting the two normal modes of motion in the trap. Phase space diagrams of the kick sequence are shown in figure \ref{fig:PhaseSpacePicture}. for the two different modes (center-of-mass and relative), both in the non-rotating frame and in the rotating frame. We can determine the evolution of a coherent state $|\alpha\rangle$ subjected to the kicks described above. For simplicity in this example, we will treat the ground state $\alpha=0$.

At the end of the simple pulse sequence, the state of the ions in a normal mode of frequency $\omega$ is:
\begin{equation}
e^{i \eta^2 (-4 \sin(\omega T)+\sin(2 \omega T))} |i \eta (1+(-2 + e^{-i \omega T})e^{-i \omega T})\rangle\label{eq:state}
\end{equation}
The phase for a given mode is given by:
\begin{equation}
\phi \approx -\frac{2 \Delta k^2 T}{m}\left(1+\frac{\omega^2 T^2}{3}\right)\label{eq:PhiModes}
\end{equation}
The phase difference between the two modes is thus given by:
\begin{align}
\Delta\phi &= -\frac{2 \Delta k^2 T^3}{3m}\left(\omega_\text{R}^2-\omega_\text{C}^2\right) \\
&=\frac{4 e^2\Delta k^2 T^3}{3d^3m^2} \label{eq:DeltaPhi2}
\end{align}
where $\omega_C$ and $\omega_R$ are the frequencies for the center of mass and relative motion modes. This is the same expression found using the Coulomb picture in equation \ref{eq:DeltaPhi}.

The phase difference in equation \ref{eq:DeltaPhi} can also be extracted by examining the phase space area enclosed by this sequence. The trajectories in the rotating frame are shown in figure \ref{fig:PhaseSpacePicture}(c) and (d). In the rotating frame all paths are driven, which leads to phase accumulation. If a coherent state is driven through a trajectory which encloses an area A in the rotating frame phase space, that coherent state acquires a phase 2A\cite{Luis,Wang}. This fact allows us to determine the phase acquired simply by calculating the area enclosed in figures \ref{fig:PhaseSpacePicture}(c) and (d). This calculation once again matches the phase in equation \ref{eq:PhiModes}.

It is worth pointing out that although the simple example illustrated in figure \ref{fig:CoulombPicture} uses the transverse modes of motion, such a phase gate also works with the axial modes of motion.  Moreover, if the axial modes of motion are used, the displacement $\delta$ is directly along the line separating the two ions, resulting in a larger modification of the Coulomb interaction. Equation \ref{eq:DeltaPhi2} applies equally for axial or transverse modes.  For transverse modes, the term in parentheses is $\omega_z^2$, while for axial modes it is $2 \omega_z^2$. So if all other parameters are held constant there is a factor of 2 greater phase when using axial modes instead of transverse. However, there is added flexibility in using transverse modes, as will be discussed below.

Unfortunately, this simple sequence of kicks has two serious limitations. First, the phase obtained from this sequence is small. Plugging realistic experimental parameters ($d=5 \mu$m, T$=100$ ns, $\Delta k =2 \times (\frac{2 \pi}{355 nm})$ into equation \ref{eq:DeltaPhi} we find a phase difference of $\pi /780$, significantly smaller than the $\pi/2$ needed for a maximally entangling phase gate. Second, the motion does not factor completely at the end of the pulse sequence, but some residual entanglement remains. This is clearly seen in equation \ref{eq:state} where the final state now depends on $\eta$, $\omega$, and $T$.  Both of these limitations can, in principle, be overcome by using more complicated pulse sequences with many laser pulses strung together to give a larger momentum kick.

The theory proposals in \cite{Duan} and \cite{GarciaRipoll} both go beyond the simple pulse sequence presented above.  In \cite{Duan}, Duan solves these problems by using many pulses in quick succession. Moreover he shows that with more complicated pulse sequences the errors can be reduced while still completing the gate in a time much faster than the trap period. This allows the scheme to be used on a pair of adjacent ions in a long chain.  If the gate is sufficiently fast, the other ions are not disturbed and the gate is scalable to large ion crystals. Unfortunately this scheme relies on a very large number of pulses ($>1000$) in a very short period of time ($< 5$ ns) and there is not currently a commercial laser available with high enough power and fast enough repetition rate to implement this scheme in our system.

In \cite{GarciaRipoll}, the trap evolution is used to control the trajectory in phase space. By correctly choosing the timing of a series of spin-dependent kicks, the relative phase accumulated by the two normal modes can be controlled and both phase space trajectories can be closed, returning the ions to their original position. Here we will present an experimentally achievable extension of their scheme with the goal of performing an entangling phase gate on two ions in less than  $1.5 \mu s$.

\begin{figure}%
\includegraphics[width=\columnwidth]{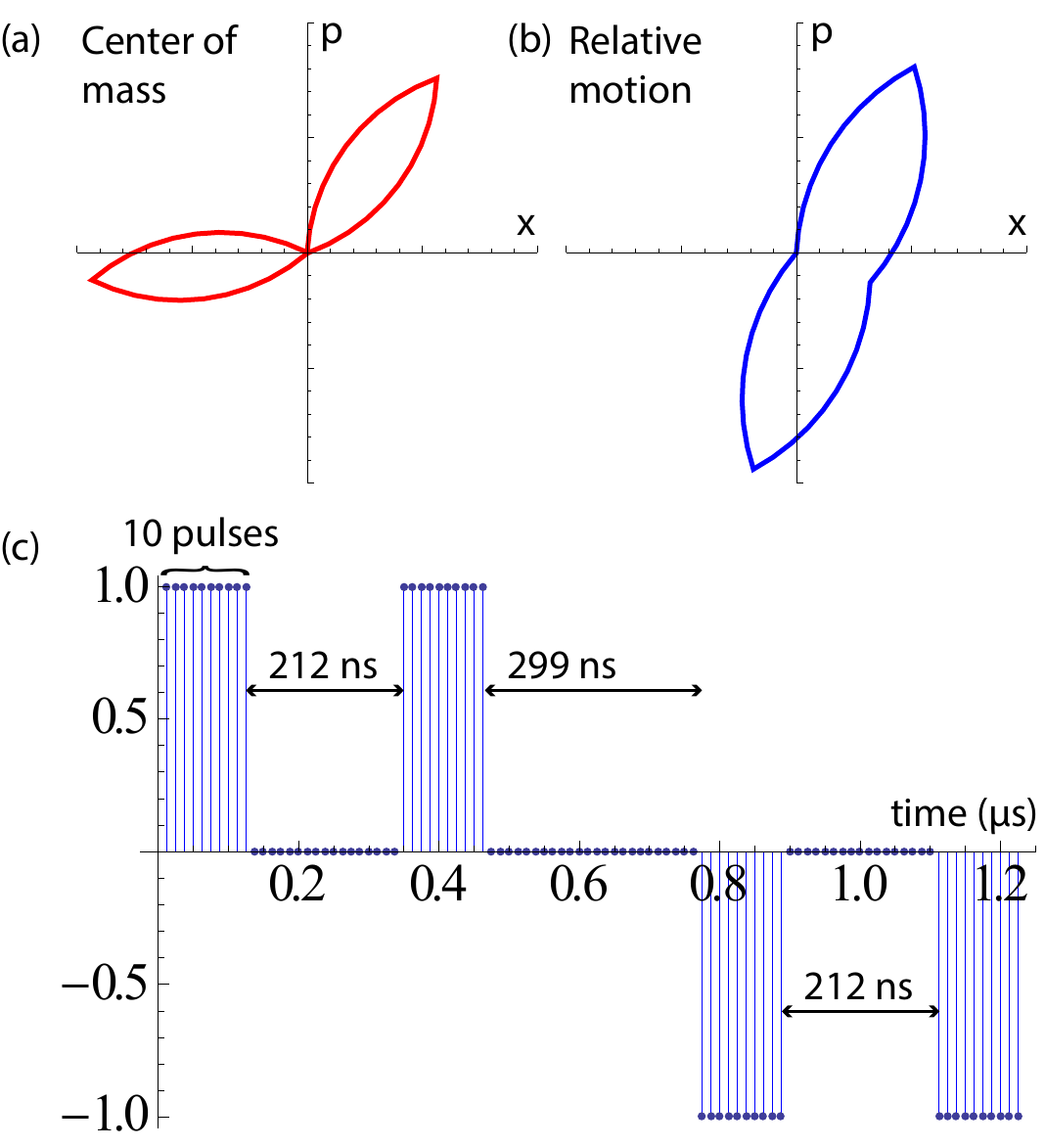}%
\caption{Phase space picture of an experimentally realizable phase gate.  (a)center of mass mode, (b) relative motion mode. In the rotating frame the direction of the spin-dependent kick rotates at the normal mode frequency. (c) Depiction of kick sequence. The ion is kicked 10 times by ten successive laser pulses with 12 ns of trap evolution between each kick.  The ion then evolves in the trap for $t_1=212$ ns then kicked 10 more times.  After a wait of $t_2=299$ ns the sequence is reversed with 10 kicks in the opposite direction, free evolution for $t_1$ and a final 10 kicks to return the ions to their original position.}%
\label{fig:realgate}%
\end{figure}

For simplicity we choose a scheme similar to that in \cite{GarciaRipoll}, but to accumulate more phase we replace each of the four spin-dependent-kicks in \cite{GarciaRipoll} with 10 spin-dependent kicks.  Experimentally each kick is derived from a single pulse of a mode locked laser with a repetition rate of 80.16 MHz, so the delay between successive kicks is 12.5 ns, which is not negligible compared to the trap period of 676 ns (frequency of 1.48 MHz).  As a result, the trap evolution between the kicks is important and must be taken into account.  We apply 10 spin-dependent kicks with 10 successive pulses from the laser.  Because each kick also flips the spin of the ion, the direction of the spin-dependent kick must be reversed between successive pulses to continue to add momentum to the system.  After the 10 spin-dependent kicks, the ion is allowed to evolve in the trap for a time $t_1=212 ns$, and then 10 more spin-dependent kicks are applied in the same direction.  The system evolves freely for a time $t_2=299$ ns, and then the first three steps are reversed, 10 more kicks in the opposite direction, evolve for $t_1$, 10 more kicks to return the system to its original location. The total gate time is $1.22 \mu s$. Figure \ref{fig:realgate} shows the path in phase space for both the center of mass and relative modes.

The scheme presented in figure \ref{fig:realgate} is just one of many possible ways to perform this phase gate. There are three constraints.  Both the normal modes must close phase space, and the differential phase must be $\pi/2$.  But even if we restrict ourselves to pulses spaced by the repetition rate of the laser during the duration of our gate there are still 98 pulses that can be used to satisfy these three conditions. Each pulse from the laser can give no momentum kick or a momentum kick of $\eta$ in either direction.  This means that there are $3^{98} = 5.7\times10^{46}$ different possible pulse sequences.  Most of those do not fulfill the constraints above, but a more detailed search is sure to reveal many solutions, it is very likely there is an optimized solution with a shorter total gate time. It is also important to note that by choosing to use the transverse instead of the axial modes of motion we can control the relative frequencies of the two normal modes which gives another way to control the phase of the gate and ensure the closure of both phase space loops \cite{Zhu2}.

\section{Conclusion}

We have demonstrated that mode-locked lasers are an extremely versatile tool in the coherent control and entanglement of trapped ions in both the fast and slow regimes.

In the slow regime, we have shown that the spectral features of the frequency comb can be used in much the same way as CW lasers, where ion-ion entanglement is produced by addressing sideband transitions.  The advantages in this regime are two fold: First, the available power enables operating much further from resonance, which reduces laser induced decoherence.  second, the broad spectrum allows direct coupling of the qubit levels using a single beam, without the experimental difficulties associated with creating a microwave beatnote between two CW beams.

In the fast regime, we have shown that it is possible to drive arbitrary rotations of a trapped ion in tens of picoseconds, which is many orders of magnitude faster than the coherence time.  We have also shown the ability to perform fast spin-dependent kicks, which opens the door to performing very fast gates. The advantage of these gates is their insensitivity to temperature, their extreme speed, and their potential for scalability.

\begin{acknowledgments}
This work is supported by grants from the U.S. Army Research Office with funding from the DARPA OLE program, IARPA, and the MURI program; the NSF PIF Program; the NSF Physics Frontier Center at JQI; and the European Commission AQUTE program.
\end{acknowledgments}

\appendix
\section{\label{sec:SechBessel} Motional Evolution Operator with Non-Zero Pulse Duration}
In section \ref{sec:SpinMotion}, equation \ref{eq:UMotBessel} was derived by approximating the pulse as a $\delta$-function.  This section examines the validity of that approximation.  The pulse duration is of order 10 ps, meaning it is several orders of magnitude faster than the trap frequency or the AOM frequency.  Therefore, the Rosen-Zener solution in section \ref{sec:SpinControl} can be used, with $\theta\rightarrow\theta\sin\left(\Delta k \hat{x} + \phi\right)$ in equations \ref{eq:RosenZenerA} and \ref{eq:RosenZenerB}:
\begin{align}
A &= \frac{\Gamma^2\left(\xi\right)}{\Gamma\left(\xi-\frac{\theta}{2\pi}\sin\left(\Delta k \hat{x} + \phi\right)\right)\Gamma\left(\xi+\frac{\theta}{2\pi}\sin\left(\Delta k \hat{x} + \phi\right)\right)}\label{eq:RosenZenerAsech} \\
B &= -\sin\left(\frac{\theta}{2}\sin\left(\Delta k \hat{x} + \phi\right)\right)\sech\left(\frac{\omega_q T_p}{2}\right)\label{eq:RosenZenerBsech}
\end{align}
The $\sx$ term in part of equation \ref{eq:RosenZenerU} is given by $iB$, which can be expanded using the Jacobi-Anger expansion as:
\begin{equation}
iB = \sech\left(\frac{\omega_q T_p}{2}\right)\sum_{\text{odd }n = -\infty}^{\infty} e^{in\phi} J_n(\theta)D\left[in\eta\right]
\label{eq:RosenZenerBSech}
\end{equation}
This is nearly identical to the $\sx$ term in equation \ref{eq:UMotBessel}, but with an overall $\sech\left(\omega_q T_p/2\right)$ term modifying the populations. The even order diffraction terms are of order $\theta^2$ or higher, which were assumed to be negligible in section \ref{sec:SpinMotion}.  Non-zero pulse duration can thus be accounted for by replacing $\theta\rightarrow\theta\,\sech\left(\omega_q T_p/2\right)$.  This will correspond to a slight reduction in the effective pulse area as compared to a $\delta$-function pulse.

\bibliography{LongUltrafastPaperRefs}

\end{document}